\documentclass[acmlarge]{acmart}

\makeatletter
\newcommand{\myconfshort}{\acmConference@shortname}
\newcommand{\myconffull}{\acmConference@name}
\newcommand{\myconfdate}{\acmConference@date}
\newcommand{\myconfloc}{\acmConference@venue}
\AtBeginDocument{
  \fancypagestyle{firstpagestyle}{
    \fancyhead{}%
    \fancyfoot[C]{}%
  }
  \fancyhf{}
  \fancyhead[LO]{\@headfootfont\shorttitle}%
  \fancyhead[RE]{\@headfootfont\@shortauthors}%
  \fancyhead[LE]{\@headfootfont\footnotesize \myconfshort, \myconfdate, \myconfloc}%
  \fancyhead[RO]{\@headfootfont\footnotesize \myconfshort, \myconfdate, \myconfloc}%
  \fancyfoot[C]{}%
}
\makeatother
\acmBooktitle{\conffull\@ (\confshort), \confdate, \confloc}

\AtBeginDocument{%
  }

\usepackage{tabularx}
\usepackage{booktabs} 
\usepackage{placeins}
\usepackage{makecell}
\usepackage{tcolorbox}
\tcbuselibrary{most}

\usepackage{graphicx}

\copyrightyear{2026}
\acmYear{2026}
\setcopyright{cc}
\setcctype{by}
\acmConference[FAccT '26]{The 2026 ACM Conference on Fairness, Accountability, and Transparency}{June 25--28, 2026}{Montreal, QC, Canada}
\acmBooktitle{The 2026 ACM Conference on Fairness, Accountability, and Transparency (FAccT '26), June 25--28, 2026, Montreal, QC, Canada}
\acmDOI{10.1145/3805689.3812253}
\acmISBN{979-8-4007-2596-8/2026/06}

\newcommand{\rev}[1]{\textcolor{black}{#1}}

\newcommand{\ts}[1]{\textcolor{black!60}{[#1]}}
\newcommand{\speaker}[2]{\textbf{\textcolor{#1}{#2}}}
\newcommand{\userline}[3]{\ts{#1} \speaker{black}{#2:} #3\\}
\newcommand{\facline}[2]{\ts{#1} \speaker{teal!60!black}{Facilitator:} #2\\}
\newcommand{\sysline}[2]{\ts{#1} \speaker{gray!70!black}{System:} \textit{#2}\\}

\begin{document}
\title[Real-Time Group Dynamics with LLM Facilitation]{Real-Time Group Dynamics with LLM Facilitation: Evidence from a Charity Allocation Task}



\author{Aaron Parisi}
\affiliation{%
  \institution{Google DeepMind}
  \country{USA}
  }
\email{aarontp@google.com}

\author{Nithum Thain}
\affiliation{%
  \institution{Google DeepMind}
  \country{USA}
  }
\email{nthain@google.com}

\author{Alden Hallak}
\affiliation{%
  \institution{Google DeepMind}
  \country{USA}
  }
\email{hallak@google.com}

\author{Vivian Tsai}
\affiliation{%
  \institution{Google DeepMind}
  \country{USA}
  }
\email{vivtsai@google.com}

\author{Crystal Qian}
\affiliation{%
  \institution{Google DeepMind}
  \country{USA}
}
\email{cjqian@google.com}


\begin{abstract}
As large language models (LLMs) evolve from single-user assistants to active participants in civic and workplace deliberation, evaluating their effects on collective decision making becomes a governance challenge. We present two empirical studies (N=879) of real-time, text-based group deliberation in an incentive-compatible charity allocation task with real financial stakes (\$7{,}200 USD). Groups of three allocate a donation budget under varying LLM facilitation conditions: Study~1 (N=204) compares three frontier models; Study~2 (N=675) compares facilitator strategies against a no-facilitation baseline. Across both studies, LLM facilitation did not significantly improve group consensus in either study, yet participants consistently preferred facilitated discussion. We additionally identify two governance-relevant risks. First, \textit{algorithmic steering}: facilitators shifted select charity-level allocations by up to 5.5 percentage points---directly affecting the final charitable payout---even when aggregate agreement metrics remained unchanged. Second, an \textit{illusion of inclusion}: participants cited inclusivity as their primary reason for preferring LLM facilitators, yet neither survey nor transcript-based measures of participation equity improved. Notably, participants reported greater trust in the process under the same
conditions where facilitators exerted directional influence on
outcomes. Together, these findings show that in AI-mediated group deliberation, perceived procedural improvement can coexist with measurable steering and unchanged participation inequality, motivating evaluation practices that treat collective outcomes, interaction dynamics, and participant perceptions as distinct governance targets.
\end{abstract}



\maketitle

\section{Introduction} 

Large language models (LLMs) are increasingly used in social and collective settings to summarize discussions, coordinate decisions, and facilitate deliberation in civic and workplace contexts~\citep{doi:10.1126/science.adq2852,Li_2025,Konya_2025,wethepeople250}. This shift raises governance-relevant challenges that are largely absent in one-on-one interactions: in group settings, values and preferences can diverge across participants, and facilitation may affect not only whether groups converge, but whose views are surfaced, how disagreement is handled, and whether the process feels legitimate.

Prior work has introduced LLM-based facilitation and mediation strategies, but evaluations often emphasize either downstream outcomes or subjective experience, without jointly examining conversational dynamics~\citep{Bu_inca_2020,tan2024robotsmiddleevaluatingllms,qian2025strategictradeoffshumansai}. \rev{Recent improvements in model latency and conversational capabilities have made it possible to evaluate LLM facilitation in real-time group discussions at scale}. This enables the study of synchronous, time-boxed interactions---scenarios such as team meetings or community town halls---that move beyond LLMs' proven capabilities in post-hoc summarization and asynchronous mediation.

We contribute two empirical studies ($N=879$) of real-time, text-based group deliberation in an incentive-compatible charity allocation task. In groups of three, participants help to allocate a real donation budget (total \$7200 USD) across nine charities. Each group completes three rounds of deliberation, with a different facilitator treatment presented in randomized order. In Study~1 (N=204), we vary the \emph{facilitator model}, comparing three frontier models (\texttt{Gemini~2.5~Flash, Claude~4.5~Haiku, and GPT-5 mini}) under a constant, minimally specified prompt. In Study~2 (N=675), we hold the model fixed (Gemini) and vary the \emph{facilitator strategy}, comparing a summarization-centric facilitator and a principles-guided facilitator against a no-facilitator baseline. Ultimately, this design enables a multi-dimensional evaluation of real-time group-AI interaction, pairing actual financial stakes with granular measures of dialogue, consensus, and multi-user preferences.


 \begin{figure*}[t]
  \centering
  \includegraphics[width=.9\textwidth,]{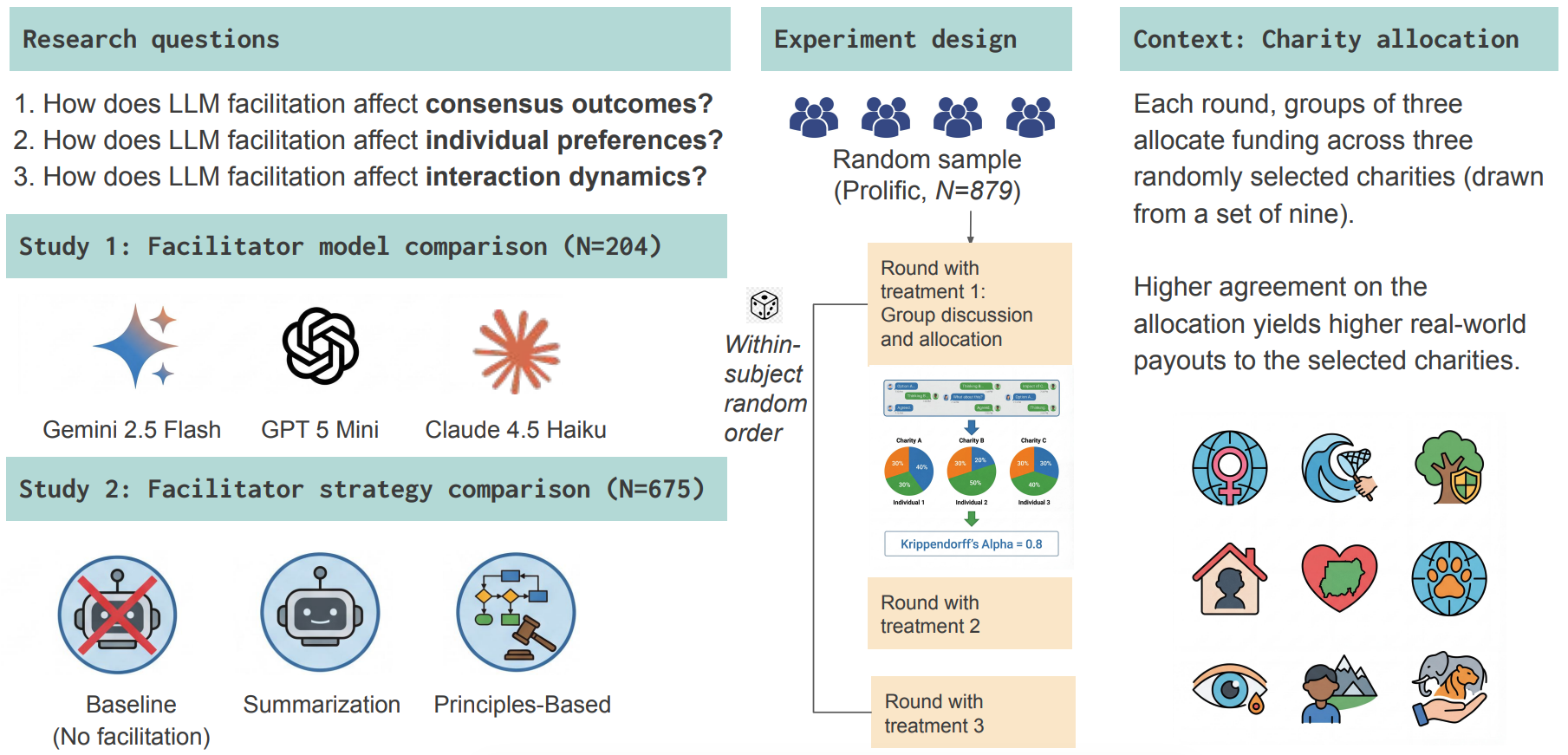}
  \caption{Experiment design overview. Participants, in groups of three, complete three rounds involving randomized access to LLM facilitation. In each round, participants discuss and allocate funds across three charities. Study 1 compares different models (N=204); Study 2 evaluates facilitation strategies against a no-facilitation control (N=675).}
 \Description{Graphical overview of experimental design, which is explicitly textually described in section \ref{sec:framework}}
  \label{fig:hero}
\end{figure*}
Our primary outcome is group-level consensus; we also evaluate how facilitation shapes decision distributions, conversational dynamics, and participant perceptions. Across both studies, facilitation by LLM agents does not significantly improve aggregate consensus, yet participants prefer facilitation over the no-facilitator baseline. At the same time, facilitators shift \emph{what} groups decide: we observe topic-specific steering in allocations even when aggregate consensus is unchanged. Participants report greater perceived inclusion and coordination with LLMs, though transcript-based measures show largely unchanged participation inequality and a slight reduction in human discussion. Finally, facilitator preference is participant- and context-dependent: participants who report higher investment in the outcomes prefer facilitation at higher rates.

Together, this work demonstrates that LLM facilitation in real-time decision making cannot be evaluated through a single outcome measure alone. We find that while real-time, LLM-based facilitation doesn't provide material improvements to group consensus, participants consistently perceive the facilitated process as better. Furthermore, facilitation can fundamentally alter how decisions are made and distributed. \rev{These findings motivate evaluation and deployment practices for multi-user AI systems that treat outcomes, processes, and perceptions as distinct governance targets.}

\subsection{Related Work}
As LLMs are increasingly used in multi-user contexts in roles such as facilitators, mediators, and moderators, it is important to evaluate how they can affect group dynamics and collective decision-making. 

\paragraph{LLM facilitation and evaluation in applied settings.} Prior work shows that LLMs as facilitators can affect group outcomes in settings such as debate, mediation, and coordination~\citep{columbiauniversity2025,Konya_2025,Li_2025,qian-etal-2025-mask}, \rev{but these evaluations often emphasize either downstream outcomes, such as agreement or task completion, or perceptual outcomes, such as satisfaction and preference~\citep{columbiasway2025}}. The literature on human-computer interaction shows that user satisfaction can be a poor proxy for downstream reasoning quality or task performance~\citep{Bu_inca_2020,tlf2024}, and finds that LLM facilitation may not reliably improve group performance~\citep{alsobay2025bringing}. \rev{We build upon this by evaluating LLM facilitation across multiple dimensions, including a primary consensus outcome, and secondary conversational dynamics, decision distributions, and participant perceptions.}

\paragraph{Comparisons against human baselines.} 

Studies comparing LLM vs. human facilitators show that participants may rate LLM-generated interventions favorably on measures such as perceived performance, information quality, or understandability~\citep{Ko_ak_2025,tan2024robotsmiddleevaluatingllms,doi:10.4258/hir.2025.31.2.200}.  \rev{However, these studies evaluate isolated facilitator utterances, constrained turn-taking, or domain-specific disputes rather than free-form group deliberation with downstream task outcomes~\citep{tan2024robotsmiddleevaluatingllms}}. \rev{By comparing multiple LLM facilitation regimes in live, synchronous group interaction, we evaluate not only participant judgments, but also consensus change, allocation shifts, and conversational dynamics.}



\paragraph{Facilitation strategies.}
Prior work on LLM-based facilitation include summarization-based mediation~\citep{doi:10.1126/science.adq2852}, principles-guided prompting, consensus-oriented refinement, and adversarial or devil's-advocate interventions~\citep{pitre-etal-2025-consensagent,chen2025multiagentconsensusseekinglarge,10.1145/3640543.3645199,10.1145/3613904.3642322,10.1007/978-981-96-0119-6_21,Papachristou_2025}. \rev{These findings suggest that facilitation strategy matters, particularly in asynchronous coordination and simulated settings. Leveraging advances in LLMs' capabilities for synchronous communication, we now have the opportunity to evaluate these facilitation methods in real-time experimental conditions.}

\paragraph{\rev{Failure modes in facilitation and group dynamics}.} Literature on generative models identifies failure modes in generative facilitation, including hallucination, overconfidence, sycophancy, and value-laden ``neutrality''~\citep{choi2025peopleagreeithink,Taubenfeld_2024,garcia2024moralturingtestevaluating,kasirzadeh2022conversationartificialintelligencealigning}. Constitutional and principle-based methods have been proposed to mitigate these risks by iteratively refining system behavior~\citep{bai2022constitutional,petridis2024constitutionalexperts,maiya2025opencharactertrainingshaping,thakur2025personasparametersfinetuningsmall}.
 \rev{Separate literature on group decision-making identifies recurring process failures in coordination tasks~\citep{janssens2022collective, karpowitz2012gender}, including premature consensus~\citep{janis1982groupthink}, biased information sharing~\citep{stasser1985pooling}, and unequal participation~\citep{woolley2010collective, siu2017deliberation}, which may be exacerbated in online settings~\citep{small2023opportunitiesrisksllmsscalable}. Guided by this literature, we develop a principles-based facilitator to address such breakdowns: model-side risks such as overconfidence and value-laden neutrality, and interaction-level failures such as weak justification, self-contained reasoning, and premature consensus.}

\paragraph{Empirical gap.} 

Taken together, prior work provides a strong foundation for understanding LLM mediation in simulated environments, asynchronous tasks, and primarily through the lens of individual user satisfaction. \rev{Building upon this foundation, our paper studies real-time, LLM-based facilitation in synchronous human group deliberation, leveraging low-latency frontier models that make live facilitation increasingly feasible at scale. This allows us to evaluate not only whether facilitation is preferred, but how it shapes agreement, decision distributions, participation, and perceived inclusion as the interaction unfolds.} \rev{Our contribution is synthesizing these outcome, process, and perception measures together in a controlled human study of live AI-mediated deliberation.}

\subsection{Contributions}

We contribute a large-scale, empirical study of LLMs as facilitation agents, presenting:
\begin{itemize}
\item \textbf{A large-scale, incentive-compatible study of LLM facilitation in a group allocation task} (N=879), \rev{evaluating a primary outcome of change in \textbf{group consensus} ($\Delta\alpha$), and secondary measures of (i) \textbf{decision distributions} (topic-level allocation shifts), (ii) \textbf{interaction dynamics} (turn-taking, participation inequality), and (iii) \textbf{participant evaluations} (preference and perceived inclusion).}
\item \textbf{Evidence of a perception--outcome divergence in multi-user settings:} participants prefer LLM facilitation over no facilitation, despite no significant gains in aggregate consensus. Participants \textit{qualitatively} felt more included with LLM facilitation, despite no quantitative improvement in participation. 
\item \textbf{Evidence of algorithmic steering in group resource allocation:} LLM facilitators can exert topic-specific steering and measurable shifts in discussion structure and resource allocation decisions, even when aggregate consensus measures remain unchanged.
\item \textbf{A comparative evaluation of model- and strategy-level facilitation designs}, motivating AI-mediated group evaluation practices that treat outcomes, process, and perception as distinct governance targets.
\end{itemize}

\section{Research Questions}
To characterize how LLM facilitation affects group decision-making, we study the following:

\begin{itemize}
    \item \textbf{RQ1 (Consensus outcomes):} How does LLM facilitation affect groups’ consensus formation on a collective allocation task?
    \item \textbf{RQ2 (Individual preferences):} How do participants evaluate different facilitation styles and models, independent of outcome improvements?
    \item \textbf{RQ3 (Interaction dynamics):} How does LLM facilitation shape conversational structure and participation patterns in group deliberation?
\end{itemize}
In addition to these pre-specified research questions, we report exploratory analyses of \textit{decision distributions} across charities, which revealed evidence of algorithmic steering under facilitation.

\section{Experiment design}\label{sec:framework}

We chose \textit{charity allocations} as the domain for the consensus-seeking task; in the task, participants help to allocate a real \$7{,}200 donation budget across nine charities spanning diverse causes (e.g., humanitarian relief, crisis response, global health, and environmental conservation).\footnote{A list of charities with inclusion criteria is provided in Appendix~\ref{app:charities}; final payouts to charities are reported in Appendix~\ref{app:charity-payouts}.} The task mirrors classic public-goods and applied budgeting settings in which groups must negotiate trade-offs across competing public priorities with limited resources \citep{Ledyard1995PublicGoodsSurvey, FehrGachter2000PunishmentPublicGoods, DellaVignaListMalmendier2012SocialPressure}, while the diverse bundle of charities elicits heterogeneous preferences across participants.

Our design introduces an incentive-compatible, real-stakes mechanism that ties outcomes to group agreement, consistent with prior studies in behavioral economics that use consequential external payouts to motivate careful decision-making~\citep{lee2021}. Participants were told that their groups would be ranked by a consensus score, and groups achieving higher consensus would receive greater weight in determining the final donation split. Concretely, we computed realized donations by aggregating groups’ final allocation vectors weighted by their consensus scores.\footnote{Instructions explaining the incentives mechanism to participants are provided in Appendix~\ref{app:participant-instructions}.}

\subsection{\rev{Experiment notation and} consensus measures}\label{sec:consensus_measures}

In the studies, participants are randomly assigned to groups of three and complete three rounds of deliberation.
In each round $i\in\{1,2,3\}$, the group allocates a unit budget across three charities
$C_i=\{c_{i0},c_{i1},c_{i2}\}$ sampled without replacement from the set of nine. Each round involves 1) an individual, initial pre-discussion
allocation across the three charities, 2) a 5-minute group discussion with team members, and 3) an individual, post-discussion allocation (stages shown in Appendix Figure~\ref{fig:round_stages}).

\paragraph{Allocations.}
Let $s\in\{0,1\}$ denote the pre-discussion (0) or post-discussion (1) stage.
In each allocation, participants distribute a \emph{unit budget} normalized to 1 across the three charities in the round; $a_{p,s,i}(c_{ij})\in[0,1]$ denotes the share of a unit budget (summing to 1) that participant $p$ allocates to charity $c_{ij}$ in round $i$ at stage $s$. We denote all of a participant p's allocations as the vector $A_{p,s,i}$.

\paragraph{Consensus.}
For each round $i$ and stage $s$, we compute Krippendorff's $\alpha$ (interval metric) \citep{Krippendorff2004}
across the three participants as the consensus score $\alpha_{s,i}$, treating each charity $j\in\{0,1,2\}$ as a unit and the
assigned share $a_{p,s,i}(c_{ij})$ as the interval-valued judgment.\footnote{\rev{For robustness, alternative consensus and influence metrics are reported in Appendix~\ref{app:alternative-metrics}, including Euclidean distance and standard deviation.}} For ease of interpretation, all $\alpha$ scores are scaled and shifted by $\alpha = (3 \cdot 100\alpha + 100)/4$, and clamped to the range [0, 100], to represent a 0–100 metric.

The change in group-level consensus score pre- and post- discussion is then $\Delta \alpha_i  = \alpha_{1,i} - \alpha_{0,i}$. 

\rev{We treat group-level change in consensus, $\Delta \alpha_i$, as the primary outcome throughout the paper. Analyses of steering effects and conversational dynamics are secondary and, where noted, exploratory.}


\subsection{\rev{Treatment design}}

In both studies, groups complete three rounds. Each round consists of an initial allocation, a group discussion stage, and a final allocation. Each round's discussion stage is conducted under one of three facilitator treatment conditions, with condition order randomized within groups.

\textbf{Timing considerations.} Each round spans approximately 15 minutes, with the live group discussion capped at five minutes. While this short window is operationally efficient for a multi-round experiment, it also targets an important and under-explored class of "micro-deliberation" settings, such as workplace stand-up meetings or rapid committee votes. Modern work and civic settings frequently rely on these brief, synchronous interactions, yet they rarely receive structured facilitation because human intervention is cost-prohibitive at this scale. Real-time LLMs offer a scalable alternative, and this design lets us test their capacity to mitigate process failures within highly compressed decision-making windows.

\textbf{Human facilitator baseline.} Our study design focuses on no-facilitator and LLM-facilitator conditions. While a human-facilitator baseline would be informative in theory, it is largely cost-prohibitive in practice for the short-horizon interactions we study. Human professional facilitators are rarely deployed for five-minute meetings, making automated interventions a unique logistical opportunity; even a marginally effective LLM facilitator would offer a substantial improvement over the status quo of no facilitation at all. Methodologically, holding the facilitator fixed also eliminates facilitator-to-facilitator variation, allowing us to compare model and strategy effects in tightly matched conditions across a much larger sample than would be feasible with human facilitation.\footnote{\rev{These design constraints are further developed in Section~\ref{sec:limitations}.}}

\subsection{\rev{LLM facilitator design}} 

 All facilitator scaffolding, system prompts, and structured-output schemas are provided in Appendix~\ref{app:facilitator-prompts}, with model configuration details (e.g., sampling parameters) in Appendix~\ref{app:llm-resources}.

\paragraph{Study 1: Model comparison.}

Study~1 compares three frontier conversational models---Gemini~2.5~Flash, Claude~4.5~Haiku, and GPT-5 mini---holding constant the same lightweight facilitator prompt.\footnote{\rev{Model selection strategy is discussed in Appendix~\ref{llm-model-selection}; full prompts and model parameters are provided in Appendix~\ref{app:llm-resources}.}} The prompt, held constant across models, specifies the task goal (e.g., helping the group reach consensus) but does not prescribe a particular facilitation strategy, approximating a naive ``default'' facilitator prompt that a practitioner might write without a more targeted intervention policy. \rev{The goal of this study is to examine baseline facilitation dynamics across several commonly used, widely deployed conversational models, rather than to offer a comprehensive model comparison.} 

\paragraph{Study 2: Strategy comparison.}
Study~2 evaluates two \emph{facilitator strategies} while holding the model fixed (Gemini~2.5~Flash), and provides a no-facilitator baseline (humans only). The two facilitator strategies are motivated by two common design patterns in prior work and practice: (i) \emph{summarization}-centric facilitation~\citep{doi:10.1126/science.adq2852}, and (ii) \emph{principles}-based, failure-mode targeting facilitation that targets a defined list of conversation failure modes and corresponding interventions, such as off-topic drift, low engagement, or premature consensus without deliberation~\citep{doi:10.1126/science.adq2852,pitre-etal-2025-consensagent,chen2025multiagentconsensusseekinglarge}.\footnote{\rev{Appendix~\ref{app:failure_solutions} reports the mapping from observed failure modes to intervention strategies used by the principles-based facilitator.}}

\subsection{Experiment interface implementation}\label{sec:experiment-interface}
We implemented the interface for both studies using Deliberate Lab \citep{qian2025deliberatelabplatformrealtime}, a platform for real-time, multi-participant experiments. The platform supports integration of LLM participants, which we leveraged to implement the facilitators.

\begin{figure}[h]
  \centering
  \includegraphics[width=.9\linewidth,]{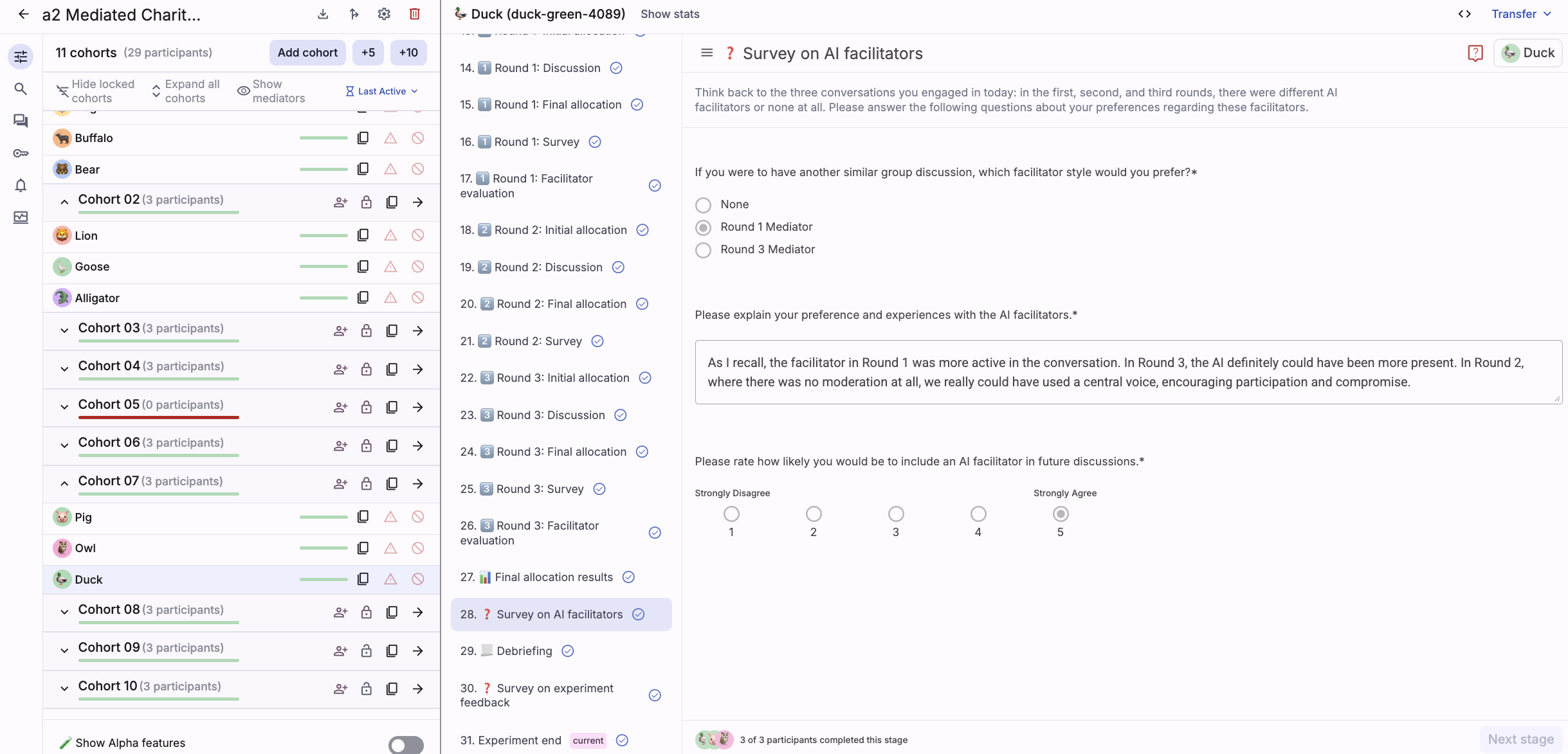}
  \caption{The Deliberate Lab experimenter interface. \textit{Left:} A list of multiple groups completing the study synchronously. \textit{Right:} A preview view of participant responses in a post-study survey. Additional images of relevant platform interfaces are provided in Appendix Figure~\ref{fig:round_stages}.}
  \Description{Overview of the DeliberateLab experimenter interface.}
  \label{fig:dl_interface}
\end{figure}

\subsection{Demographic and qualitative measures}\label{sec:qualitative_measures}

We collect aggregated demographic data via Prolific (age, gender, education;
Appendix~\ref{app:participant_demographics}) \citep{prolific2025}. Participants additionally complete three brief self-report surveys: i) an initial survey on participant backgrounds and experience using AI tools, ii) post-round surveys measuring perceptions of the given round and (when applicable) evaluations of the LLM facilitator, and iii) a post-study survey with overall reflections and comparative judgements across the three rounds. Survey items related to participants' group behaviors are adapted from the Need for Closure Scale (NFCS) \citep{WebsterKruglanski1994NFCS} and the Conflict-Handling Best-Worst Scale (CHBWS) \citep{DalyLeeSoutarRasmi2010CHBWS}, and attitudes towards AI are measured using the Technology Acceptance Model (TAM) \citep{Davis1989TAM}. Full survey instruments and results are provided in Appendix~\ref{app:survey-questions}.

\subsection{Human study recruitment}\label{sec:recruitment}
\paragraph{Recruitment.} Participants were recruited from Prolific \cite{prolific2025} under institutional IRB-approved protocols and rerouted to the Deliberate Lab platform. After providing informed consent, they were assigned to live 3-person groups with random, pseudonymous, animal-based identifiers (e.g. `Lion`, `Goose`). Using Prolific's standard sampling procedure, we recruited English-fluent participants for the full trial, excluding the 20 individuals who took part in our study pilots (N=20).\footnote{Participant demographics are provided in Appendix~\ref{app:participant_demographics}.}

\paragraph{Incentives and payment.}
Participants were compensated at \$15 USD/hour. To promote incentive compatibility, we informed participants that greater within-group consensus would increase the weight placed on their group's final allocation in determining the experiment's real charitable payout. 
Before the rounds began, a comprehension check verified that they understood the key incentive: greater agreement within the group increased the influence of their group's final allocation on the real charitable payout.

\paragraph{Sample size rationale.}
RQ1 (consensus) is defined at the group level, so the independent unit is the 3-person group.\footnote{\rev{Power calculations were conducted for the primary group-level analysis of consensus change ($\Delta_\alpha$). Preference analyses were pre-specified, but steering effects and conversational dynamics were not separately powered and should be interpreted as exploratory.}} We power paired, group-level contrasts in $\Delta\alpha$; for a paired contrast, the required number of groups is
\[
G=\left\lceil \frac{(z_{1-\alpha^\star/2}+z_{1-\beta})^2}{d^2}\right\rceil,
\]
where $d$ is the standardized effect size of the paired difference and we target $1-\beta=0.80$. Study~1 compares three models (three pairwise contrasts; $\alpha^\star=0.05/3$). Study 1 is a screening study to identify practically meaningful differences between frontier models with minimal prompt specification; we power it to detect moderate effects ($d = 0.35$). Study~2 compares two strategies
against a no-facilitator baseline (two contrasts; $\alpha^\star=0.05/2$). To evaluate how prompt-level changes might yield small but meaningful shifts in group consensus, we power Study 2 to detect small effects ($d=0.19$).
\paragraph{Participant statistics.}
Both studies were completed in November 2025.
In Study~1, we recruited $N=258$ participants; $N=204$ in 68 groups completed the full study. In Study~2, we recruited $N=796$ participants; $N=675$ in 225 groups completed the full task. If any member of a group dropped out during a round, we excluded data for the full group. Mean completion time was $51.0 \pm 16.8$ minutes in Study~1 and $52.8 \pm 17.4$ minutes in Study~2.


\section{Results}



We find no aggregate consensus gains (RQ1) relative to baseline in the LLM facilitation treatment conditions, but strong user preference for facilitation (RQ2). In RQ3, we find that introducing a facilitator can shift group dynamics. \footnote{\rev{Illustrative conversation transcripts across both studies are provided in Appendix~\ref{app:transcripts}.}}

\subsection{RQ1: Consensus outcomes}


\begin{figure}[h!]
  \centering
  \includegraphics[width=.8\linewidth,]
  {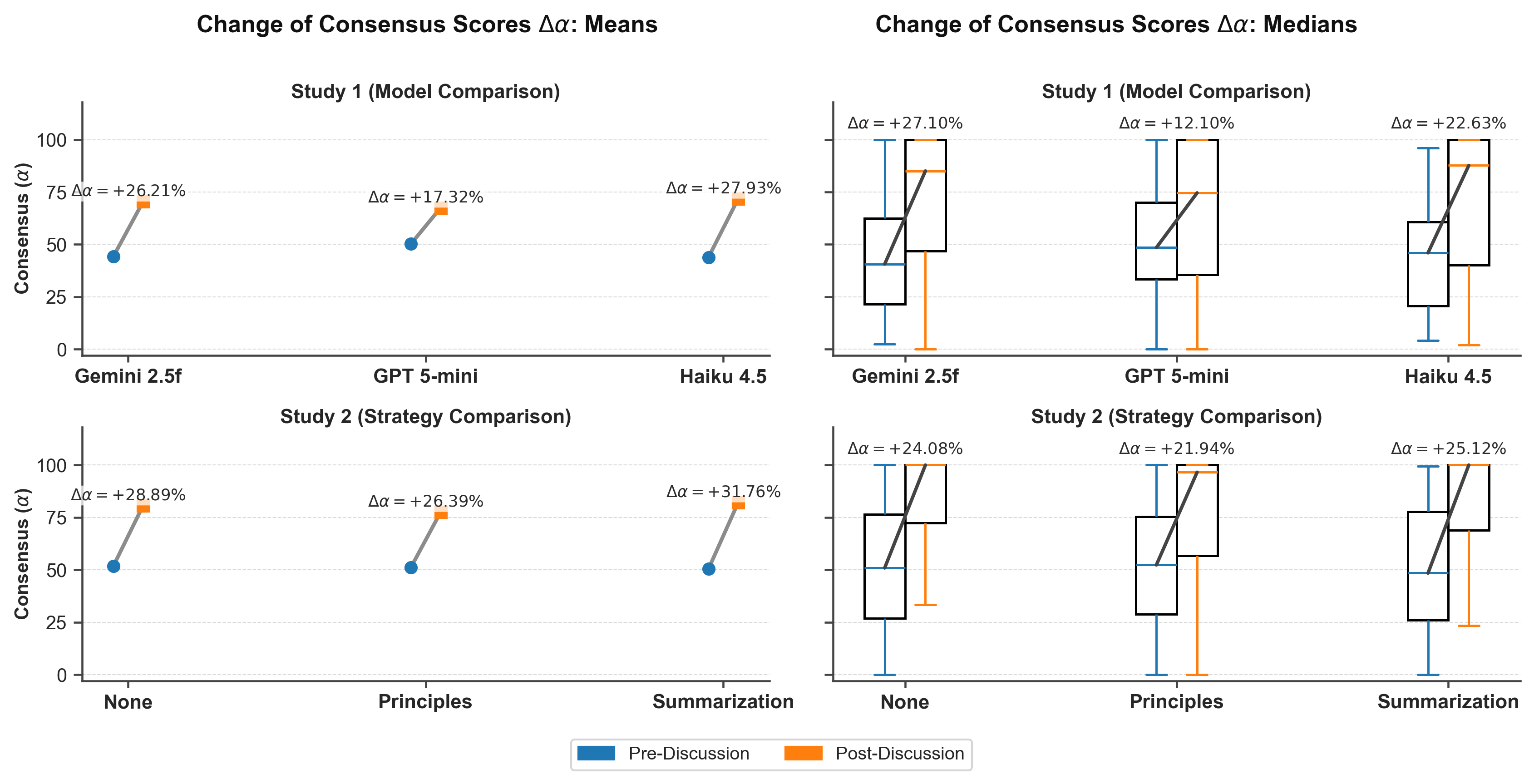}
    \caption{Changes in group consensus score across studies and rounds. \textit{Left.} Mean changes with standard error bands. \textit{Right.} Median changes with IQRs. Full values across rounds are provided in Appendix~\ref{app:consensus_scores}.}
    \Description{Change in group mean / median consensus scores. Notably, there is no measurable / visible difference in either mean / median score detectable, a fact which is discussed in detail in this section.}
      \label{fig:consensus_histograms}
\end{figure}

Across both studies, we observe no statistically significant aggregate effect of LLM facilitation on consensus relative to the unfacilitated control, although the effects of facilitation are consistently positive. It may be that this particular task affords limited room for improvement: post-conversation consensus scores were high at baseline (round one median of $\alpha_1 = 91.1$, with an aggregate IQR [68.3, 100]; see Appendix Table~\ref{tab:ordering_effects_median}), suggesting a ceiling effect and limited opportunity for participants to benefit from LLM facilitation. We observe a slight, non-confounding learning effect and increased coordination across rounds; post-discussion $\alpha_1$ increases over round number.

\begin{figure}[h!]
  \centering
  \includegraphics[width=1.0\linewidth,]{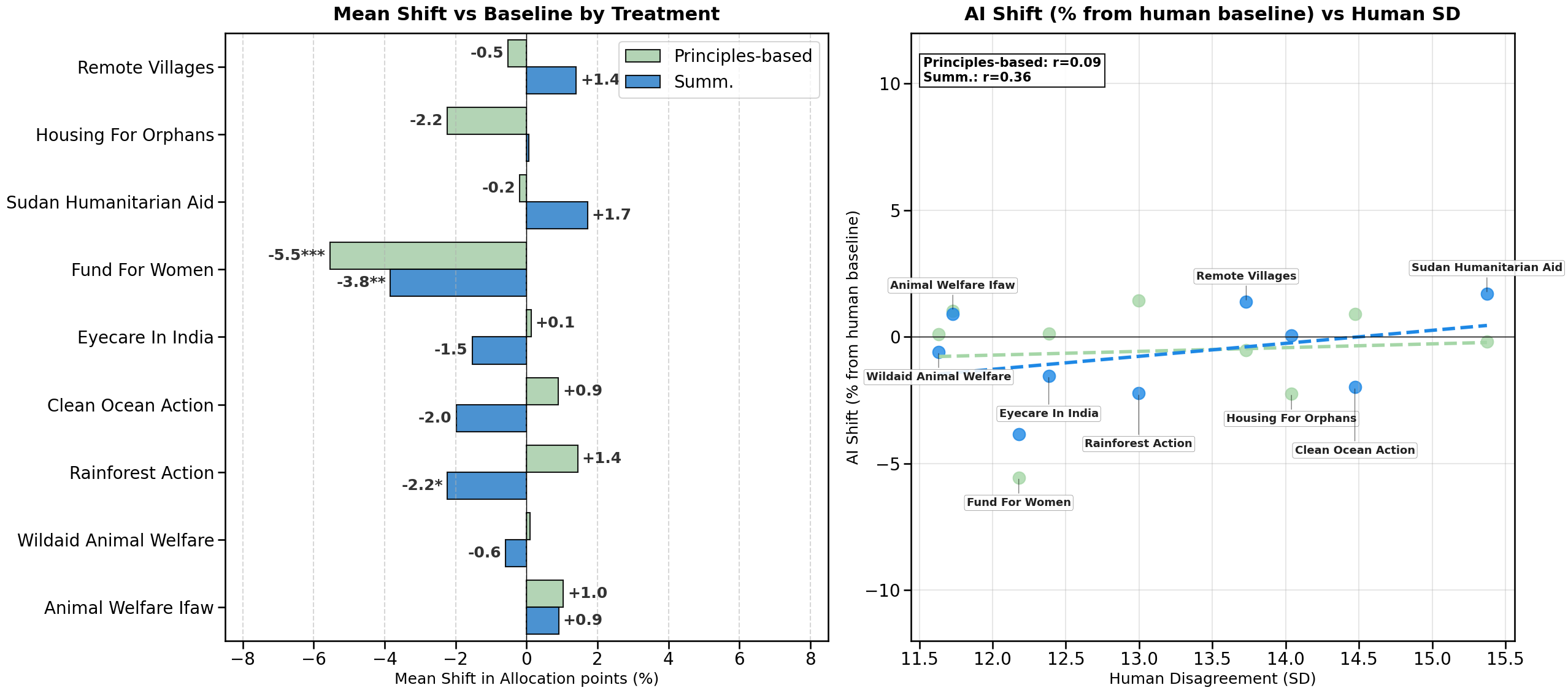}
  \caption{\textbf{Allocation steering by charity (Study 2)} \textit{Left} Bars show the change (percentage points) in the \emph{average post-discussion allocation} under each strategy-driven facilitator relative to the human-only baseline for that charity (stars denote statistically significant shifts). \textit{Right} AI shift (\%) vs the standard deviation (SD) of human baselines. }
  \Description{Two plots detailing the shift in voting practices induced by the LLM. We note significant negative shifts for the funds for women for both the principles-based and summarization treatments.}
  \label{fig:treatment_charity_effects}
\end{figure}

Despite no significant changes in aggregate consensus scores, we find evidence of \textbf{algorithmic steering}; the presence of facilitation shifts the distributions of allocations across charities. The left panel of Figure~\ref{fig:treatment_charity_effects} shows charity-specific shifts in allocations under the two strategy-driven facilitators; allocations to the \textit{Global Fund for Women} in the \textit{Principles} and \textit{Summarization} conditions exhibit $-5.5$ pp and $-3.8$ pp lower allocations relative to the no-facilitator baseline ($p<0.001$).\footnote{Significance computed using Welch's two-sided t-test.} Similarly, the Gemini 2.5 facilitator in Study 1 disproportionately \textit{increased} relative allocations to IFAW (Animal Welfare) relative to the all-human baseline (Fig.~\ref{fig:treatment_charity_effects}). Transcript analysis reveals a primary mechanism for this observed phenomenon: by summarizing and revisiting human-proposed ideas and allocations, LLM facilitators frequently repeat and formalize outlier proposals. This algorithmic steering represents a subtle accountability risk: an allegedly neutral facilitator can introduce a measurable directional influence on a real financial outcome, through simple mechanisms such as summarization. 

The right panel of Figure~\ref{fig:treatment_charity_effects} shows that the principles- and summarization-based facilitation strategies can shift the allocation downwards or upwards, depending on the charity $a_{p,s,i}(c_i)$. Why might facilitators produce larger per-charity allocation shifts for charities with higher-variance allocations? A qualitative analysis of conversation transcripts suggests one mechanism: facilitators tend to restate participant-proposed splits, which can turn outlier proposals into salient numeric anchors or increase their perceived legitimacy \citep{tversky1974judgment,galinsky2001firstoffers,mehta1994focalpoints,hovland1951credibility,hasher1977frequency,brashier2020judging}.\footnote{This is expanded upon in Appendix Figure~\ref{app:argument_analysis_women} and Figure~\ref{app:argument_analysis_ifaw}.}

\subsection{RQ2: Individual preferences}

\subsubsection{Self-reported preferences}
\begin{figure}[h!]
  \centering

  \includegraphics[width=.8\linewidth,]{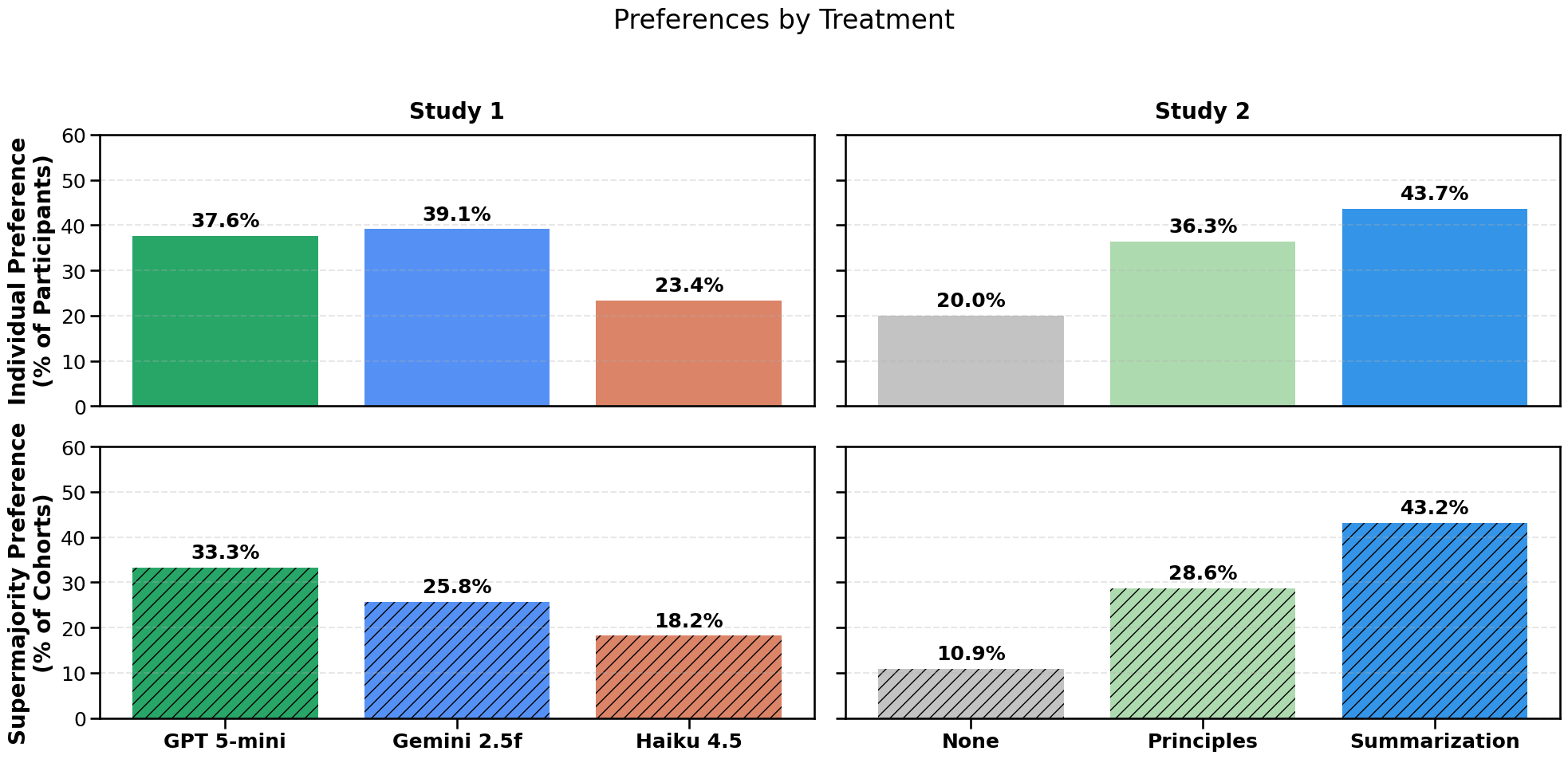}
    \caption{Participant preferences by facilitator. \textbf{Top.} Normalized individual responses show consistent preference for LLM facilitation over the unfacilitated control, with differences across facilitation styles and models. \textbf{Bottom.} When we visualize supermajority preferences ($\geq 2$ participants in group prefer the same treatment), similar trends emerge.}
   \Description{Plots of individual and supermajority (N>=2) preference for specific treatments. It is clear that there is a significant ($p<0.001$) preference for LLM-facilitators over No-mediator baselines.}
      \label{fig:fac_pref}
\end{figure}

Although facilitators do not significantly increase post-discussion consensus outcomes (RQ1), participants show a clear and statistically significant preference for LLM facilitation over the no-facilitator baseline in Study 2 ($\chi^2$ test, $p < 0.001$). Figure~\ref{fig:fac_pref} (top row) plots participant-level responses for their preferred condition in each study: participants prefer Gemini and GPT to Haiku in Study~1 ($\chi^2$ test, $p < 0.05$), and disproportionately select LLM facilitation over the human-only baseline in Study 2. These patterns persist when aggregating preferences at the group level. Figure~\ref{fig:fac_pref} (bottom row) reports the frequency with which a condition is preferred by a supermajority ($\geq2$ of 3 participants). Supermajority preferences follow similar preference patterns, indicating that preferences can be aligned within groups.

\subsubsection{Thematic analysis.}\label{sec:thematic-analysis}
To interpret facilitator preferences, we conducted a thematic analysis of free-text responses from Study~2~\citep{braun2006thematic} on preferred treatment rationale, coding aggregate themes and providing code counts below. Participants were \emph{not} told which facilitator type they received; in practice, they saw labels such as ``Facilitator 1'' and ``Facilitator 2'' rather than the descriptive condition names referenced throughout this text.

\paragraph{Summarization-preferred.} Those who preferred the summarization facilitator described it as adding \textbf{structure via recap/coordination} ($N{=}141$), e.g., ``helped by summarizing what everyone's points were'' (Participant 449) and ``summarized who was in agreement and who wasn't and why'' (P138). A second recurring theme was increasing \textbf{inclusivity} ($N{=}90$; ``made sure everyone was heard''), alongside a smaller emphasis on \textbf{maintaining focus} ($N{=}41$).

\paragraph{Principles-preferred.} The dominant theme was \textbf{productive questioning} ($N{=}134$), framed as prompting deeper reasoning and moving discussion forward: the facilitator ``asked relevant questions which prompted us to think deeper'' (P622). Participants also attributed \textbf{inclusivity} to this facilitator ($N{=}106$).

\paragraph{No-facilitator-preferred.} Responses emphasized that the group already had high \textbf{agreement} ($N{=}239$) and \textbf{outcome quality} ($N{=}112$), reporting that facilitation was unnecessary or disrupted flow (e.g., ``we can easily talk without AI\ldots it disrupts natural\ldots flow of thinking''; P511), consistent with the high consensus scores observed across this task. Notably, participants preferring this control treatment did not invoke \textbf{inclusivity} or \textbf{feeling heard} in their free-text responses, in contrast to participants who preferred the AI-facilitator conditions.

Negative feedback around the two facilitator strategies concentrated on timing and verbosity. For the summarization facilitator, the most common request was \textbf{interrupting less} ($N{=}71$; ``it just wanted me to repeat and clarify what I already said''; P537), though others wanted \textbf{more proactive guidance} ($N{=}39$), such as proposing concrete options. For the Principles facilitator, participants more often requested \textbf{more (or more targeted) questions} ($N{=}52$) while still flagging interruptions or message length as friction points ($N{=}46$).

\subsubsection{Individual differences in facilitator preference}
\begin{figure}[h!]
  \centering

  \includegraphics[width=1.0\linewidth,]{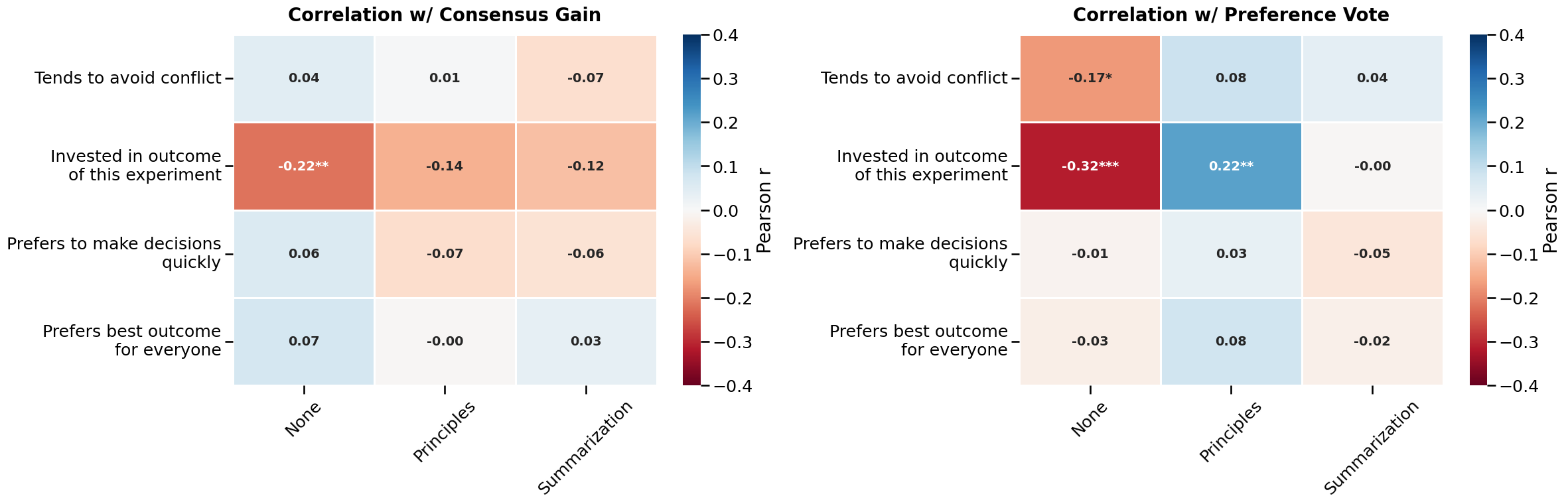}
    \caption{\textit{Left}. Pearson correlation (r) of participant's self-identified traits vs. change in group consensus outcomes. There is a slight negative correlation between participants self-identifying as being invested in the outcome, and the change in consensus score within their group. \textit{Right}. Pearson correlation (r) of participants' traits vs. preference for the human-only baseline. There is a statistically significant negative correlation between participants self-identifying as being invested in the outcome, and preferring the 'None' treatment (r=-0.32***).}
    \Description{}{Pearson's correlation of quantitative self-reported characteristics and outcomes (consensus and preference for specific models, respectively). Notably, participants who reported a strong investment in the outcome of this experiment were negatively correlated with the None facilitator in both consensus and preference, meaning their ability to reach consensus was hindered by not having a facilitator, and they strongly preferred LLM facilitators over None.}
      \label{fig:personality_perceptive}
\end{figure}

Given differing preference rationales, we explore correlations between self-reported participant traits and condition preference. In Figure~\ref{fig:personality_perceptive}, we compute Pearson correlations between self-described attributes from the pre-task survey (Likert numerics) and preferences for each treatment condition.\footnote{\rev{*p < .05, **p < .01, ***p < .001}} 

In the left panel of Figure~\ref{fig:personality_perceptive}, we observe a marginally significant \textbf{negative} correlation between self-identified investment in the task's outcome and change in group consensus score $\Delta_\alpha$: more invested participants tend to exhibit less improvement in consensus under no facilitation. In the right panel, participants who report being more invested in the outcome also appear to want more process support: they are substantially less likely to prefer the no-facilitator baseline ($r=-0.32^{***}$) and slightly more likely to prefer the \textbf{Principles-based facilitator} ($r=0.22^{**}$), which adopts a more interventionist style (questions, prompts, and nudges) than the more passive recap style of the \textbf{Summarization-based} facilitator. Taken together, these patterns suggest that invested participants are both less likely to reach consensus without facilitation and more likely to prefer active facilitation when it is available. \rev{This interpretation is also consistent with the observed behavior of the Principles-based facilitator. As shown in Appendix~\ref{app:failure_solutions}, its most frequent triggers were \textit{NoJustificationOrPrematureConsensus} ($N=636$) and \textit{SelfContainedReasoningOnly} ($N=380$), suggesting that it most often intervened when participants reached conversational states that called for more active support: eliciting justification, surfacing implicit reasoning, and encouraging engagement with others' views, rather than merely summarizing what had already been said.}


\subsection{RQ3: Interaction dynamics}

\paragraph{Conversation structure.} Table~\ref{tab:conversation_stats} reports the number of turns and total words per round for humans and LLMs. Groups engaged in substantial interaction within the five-minute discussion window. In the no-LLM baseline, groups exchanged an average of 15.0 human turns per round—roughly 3 unique messages per minute. Under facilitation, participants contributed fewer turns (approximately 10--14 per round), but this was offset by multiple facilitator interventions; the principles-based facilitator, for instance, responded about once per minute (Table~\ref{tab:conversation_stats}).

LLM facilitation increases total message volume relative to the no-LLM baseline, driven by additional LLM messages rather than greater human participation. Notably, in RQ2 we found that participants prefer Gemini and GPT facilitation over Haiku. Here, we find that Gemini and GPT are also more verbose—contributing longer messages at higher volume—which in turn leaves less room for human messages on average. Haiku, by contrast, is less verbose and its groups exhibit slightly higher human participation, yet participants preferred it the least.

Despite these shifts in volume and turn-taking, facilitation has limited effects on participation balance. We measure participation \textbf{inequality} as the variance in number of turns across the three group members within a group. Inequality in human contributions remains broadly stable across conditions, with no consistent reduction relative to the unfacilitated baseline~(Table~\ref{tab:conversation_stats}). These quantitative measures contrast with the self-reports in Section~\ref{sec:thematic-analysis}, where participants describe both facilitation strategies as improving perceived inclusivity.

\begin{table}[t]
    \centering
    \small
    \setlength{\tabcolsep}{3pt}
    \caption{\textbf{Volume of outputs by participant type and condition.} Mean outputs (turns, total words) are reported per conversation
for human and facilitator participants. Inequality is the variance in number of turns within a group. Standard
errors are in parentheses.}
    \label{tab:conversation_stats}
    \begin{tabular}{lccccccc}
        \toprule
        & \multicolumn{3}{c}{\textbf{Human Outputs (N=3)}} & \multicolumn{2}{c}{\textbf{LLM Outputs}} & \multicolumn{2}{c}{\textbf{Total}} \\
        \cmidrule(lr){2-4} \cmidrule(lr){5-6} \cmidrule(lr){7-8}
        \textbf{Condition} & \textbf{Turns} & \textbf{Words} & \textbf{Inequality} & \textbf{Turns} & \textbf{Words} & \textbf{Turns} & \textbf{Words} \\
        \midrule
        \multicolumn{8}{l}{\textbf{Study 1: Model comparison}} \\
        Claude 4.5 Haiku & \makecell{12.4 \\ (0.7)} & \makecell{122.0 \\ (7.0)} & \makecell{2.51 \\ (0.50)} & \makecell{2.0 \\ (0.3)} & \makecell{226.2 \\ (36.2)} & \makecell{14.4 \\ (0.8)} & \makecell{348.2 \\ (38.3)} \\
        Gemini 2.5 Flash & \makecell{10.3 \\ (0.7)} & \makecell{116.2 \\ (7.0)} & \makecell{1.75 \\ (0.36)} & \makecell{6.1 \\ (0.3)} & \makecell{538.5 \\ (33.0)} & \makecell{16.4 \\ (0.9)} & \makecell{654.7 \\ (37.6)} \\
        GPT-5 mini & \makecell{10.9 \\ (0.6)} & \makecell{99.5 \\ (7.6)} & \makecell{2.07 \\ (0.37)} & \makecell{5.1 \\ (0.2)} & \makecell{578.9 \\ (34.0)} & \makecell{16.0 \\ (0.7)} & \makecell{678.4 \\ (36.1)} \\
        \midrule
        \multicolumn{8}{l}{\textbf{Study 2: Strategy comparison}} \\
        None (Control) & \makecell{15.0 \\ (0.4)} & \makecell{134.4 \\ (3.8)} & \makecell{2.22 \\ (0.19)} & \makecell{0.0 \\ (0.0)} & \makecell{0.0 \\ (0.0)} & \makecell{15.0 \\ (0.4)} & \makecell{134.4 \\ (3.8)} \\
        Summarization & \makecell{14.0 \\ (0.4)} & \makecell{128.4 \\ (4.1)} & \makecell{2.39 \\ (0.25)} & \makecell{4.3 \\ (0.1)} & \makecell{186.1 \\ (6.3)} & \makecell{18.4 \\ (0.5)} & \makecell{314.5 \\ (8.5)} \\
        Principles & \makecell{14.0 \\ (0.4)} & \makecell{132.4 \\ (4.0)} & \makecell{2.14 \\ (0.23)} & \makecell{3.9 \\ (0.1)} & \makecell{119.9 \\ (4.9)} & \makecell{17.9 \\ (0.4)} & \makecell{252.3 \\ (6.7)} \\
        \bottomrule
    \end{tabular}
\end{table}

\paragraph{Language markers.} In addition to analyzing statistical measures such as volume of messages, we present trends and observations around the \textit{content} of chat transcripts. For each participant message, we conducted thematic analysis of case-insensitive whole-word instances of the following behavioral markers: agreement (\textit{agree, yes, correct, right, ok, sure, absolutely, exactly}), reasoning (\textit{because, therefore, think, believe, reason, why, since, so}), collective framing (\textit{we, us, our, group, team}), and individual framing (\textit{I, my, me, mine}). Appendix Table~\ref{tab:speech-tokens} reports average per-participant token counts for these markers. In Study~2, holding the model constant, participants used these markers at similar rates regardless of how the facilitator was prompted. In Study~1 (model comparison), different dynamics emerge. Haiku elicits stronger collective framing relative to other models per user (1.41 tokens/user, compared to 1.04 for Gemini). Gemini elicits more reasoning language ($2.69$ tokens/user, ${\sim}$2.00 for other models), and GPT elicits the shortest messages (33.2 words/user) with higher individual framing (6.37 tokens/user). This suggests that while baseline models may not significantly alter consensus outcomes, their inherent strategies may subtly influence conversational dynamics and how participants approach their tasks.  

\section{Discussion}

\paragraph{\rev{Preference and collective outcomes can diverge under facilitation.}} Across both studies, AI facilitation did not significantly improve group consensus ($\Delta\alpha$), yet participants consistently preferred facilitated discussions over unfacilitated ones. \rev{This extends single-user HCI findings that user perceptions can diverge from downstream performance~\citep{Bu_inca_2020,tlf2024}.} In multi-user settings, the divergence is more consequential: a facilitator can be experienced as useful or fair without improving collective outcomes, and different participants may value different features of the process. \rev{Many deployed LLMs are developed and selected using human preference signals~\citep{ouyang2022training,bai2022constitutional,maiya2025opencharactertrainingshaping}.} Our results suggest that in group settings, those signals reward facilitators that feel effective while leaving collective outcomes unchanged — and that this preference may partly reflect novelty rather than sustained value. Confirming whether it persists over time requires longitudinal research. For governance, this implies that high user-satisfaction scores are insufficient, and often misleading, proxies for the safety and efficacy of conversational AI systems.

\paragraph{Algorithmic steering: facilitation can shift decisions without changing consensus}

Even when overall agreement ($\Delta \alpha$) does not measurably improve, facilitation can reshape what groups ultimately decide.
We observe charity-specific shifts in allocations under LLM facilitation (e.g., reduced allocations to Global Fund for Women). Qualitatively, facilitators---particularly the summarization agents---restated participant-proposed splits, which can normalize outlier suggestions in high-variance topics. This shows that while facilitation may not affect outcome measures (e.g. $\alpha$), it can still exert directional influence. Deliberative quality depends not only on whether groups converge, but also on how decisions were surfaced, normalized, and negotiated \citep{small2023opportunitiesrisksllmsscalable}; our results provide empirical evidence that these distributional effects can arise even when a primary agreement metric remains unchanged. This raises an accountability concern: a facilitator presented as neutral can redirect real resources through its influence on dialogue, even when core metrics suggest no effect. In this experiment, the $-5.5$ percentage-point shift in Global Fund for Women allocations under the Principles facilitator translated directly into the final charitable payout (Table~\ref{tab:charity_payouts}), demonstrating that algorithmic steering has financial consequences even in a controlled study.

\paragraph{\rev{Illusion of inclusion: perceived inclusion does not imply quantitative gains to inclusivity.}} Participants frequently described facilitators as improving inclusivity, but we observe no improvement in conversational shares and a modest displacement of human turns when a facilitator is present~\citep{few2007illusion}. \rev{This echoes work on procedural justice and deliberation: people care deeply about voice, acknowledgment, and respectful treatment, yet inequality in who speaks and who is influential persists even in settings designed to be fair~\citep{tyler2003procedural,siu2017deliberation,karpowitz2012gender}.} Our findings indicate that preference-tuned LLM facilitators improve \textit{perceived} inclusiveness—through acknowledgment, summaries, and directed responses—without materially equalizing participation. This is a sociotechnical risk: facilitators can provide the conversational texture of an inclusive process while leaving the underlying distribution of voice unchanged. \rev{Future work should pair perception-based measures of inclusion with direct measures of conversational share, influence, and uptake. Combined with the steering effects reported above, this creates a compound risk: participants report greater trust in the process under exactly the conditions where facilitators exert the most directional influence on outcomes.}

\paragraph{\rev{Facilitation preferences are context-dependent.}} Preferences are not uniform: participants who report greater investment in the outcome are less likely to prefer the no-facilitator baseline and more likely to prefer the more dynamic \textit{Principles}-style facilitator. \rev{This is consistent with prior work showing that mediation style shapes both how the mediator is perceived and how much support participants want from the process~\citep{10.1145/3613904.3642322}.} Comparing facilitated and unfacilitated conditions, interventions that help under low-coordination or higher-stakes conditions can feel intrusive when groups are already aligned. Facilitation policies should be evaluated not only on average preference, but on their effectiveness across varying group dynamics.
\paragraph{\rev{Implications for evaluation and deployment.}}
Taken together, these results argue for collective evaluations that separate (i) collective outcomes (e.g., $\Delta\alpha$), (ii) decision distributions across tasks (e.g. topic-level shifts and steering), (iii) procedural dynamics (e.g. participation balance and conversational displacement), and (iv) perception (e.g. preference and felt inclusion). We caution against deploying facilitators and other sociotechnical systems based primarily on preference measures: in our study, we show that preference signals may not align with outcome measures in group settings. This is especially relevant for workplace facilitation tools and citizen deliberation platforms, where legitimacy and equity depend on procedural dynamics (who speaks, what gets discussed, and how consensus is reached); in these settings, preference-outcome mismatches can amplify idiosyncratic views, enable strategic gaming of participating LLM agents, or shift facilitator behavior toward satisfying organizers rather than the participating group~\citep{tyler2003procedural, small2023opportunitiesrisksllmsscalable}.
\rev{These findings extend prior work on AI assistance and facilitation by showing that real-time group settings require a broader evaluative lens, as facilitators may be experienced as effective while remaining imperfectly aligned with group-level outcomes and deliberative processes.}

\section{Conclusion}
As LLMs move from single-user assistance to collective settings, evaluating how they affect collective outcomes and social dynamics is critical. \rev{In collective settings, facilitators may shape not only whether groups converge, but also what they decide, how the discussion unfolds, and how the process is experienced.} We study this question in a large-scale charity-allocation task with two experiments that vary either the facilitator model (Study 1) or the facilitator strategy (Study 2), allowing us to measure consensus change, decision distributions, interaction dynamics, and participant perceptions within the same setting.

Across both studies, LLM facilitation did not produce significant changes in consensus ($\Delta\alpha$), \rev{our primary outcome}, yet participants consistently found this process preferable over the no-facilitator baseline and differentiated between models and facilitation styles. At the same time, facilitation changed the \textit{shape} of deliberation: it modestly reduced human discussion, left participation inequality largely unchanged, and produced charity-specific steering effects even when aggregate agreement remained unchanged. \rev{These findings show that in group settings, perceived procedural improvement and directional influence can coexist with null aggregate outcome effects.}

Taken together, our results emphasize that building agentic systems to improve group dynamics should not focus solely on outcome measures or preference measures. \rev{Instead, AI-mediated group dynamics should be assessed across multiple dimensions, including collective outcomes, decision distributions, procedural dynamics, and participant perceptions. This is especially important for workplace facilitation tools, deliberation platforms, and other multi-user systems, where legitimacy depends not only on what a group decides, but on how consensus is reached and whose voices shape it.} LLMs as real-time facilitators are a scalable mechanism for structuring group deliberation---one that demands evaluation, governance, and deployment practices calibrated to collective decision-making, not individual assistance.


\section{Limitations}\label{sec:limitations}
Our results are specific to a controlled charity-allocation task with small groups, short synchronous discussions, and relatively lightweight facilitator prompts. These design choices let us compare facilitation conditions cleanly, but also narrow the range of settings to which the findings directly apply.

\paragraph{Lack of a human facilitator baseline.}
We compare LLM-facilitator and no-facilitator conditions, absent a human
facilitator baseline. Accordingly, our results identify differences
across LLM conditions and relative to no facilitation, but do not establish whether
LLM facilitation reproduces, improves on, or falls short of human facilitation. We
note that the short-form deliberation settings we study (e.g., workplace standups,
rapid committee votes) rarely employ human facilitators due to cost, making LLM
facilitation a comparison against the realistic status quo of no facilitation at all.
Holding the facilitator fixed also eliminates facilitator-to-facilitator variation,
enabling tighter treatment comparisons at larger sample sizes than human facilitation
would permit. Direct human--LLM comparisons, particularly on process quality, steering,
and preference--outcome divergence, remain an important direction for future work.

\paragraph{\rev{Discussion duration and time to consensus.}}
The five-minute discussion cap targets micro-deliberation settings but precludes
measuring time-to-consensus as an independent variable and limits the depth of
deliberation groups can achieve. Future work should use longer or open-ended discussion
windows to evaluate whether the effects observed here---algorithmic steering, illusion
of inclusion, null $\Delta\alpha$---persist, attenuate, or amplify over time.
\paragraph{High baseline agreement and low-conflict structure.}
Participants had access to the same charity information, faced limited strategic
incentives, and often started from high levels of agreement (median post-discussion
$\alpha_1 = 100$). This likely constrained the room for facilitation to improve
consensus. Facilitation may have larger effects in settings with information
asymmetries, role-differentiated expertise, or genuine conflicts of interest, where
surfacing hidden constraints and reconciling competing accounts is central to the task.

\paragraph{Facilitator design.}
The facilitators in both studies were intentionally lightweight and not tuned to the
population or task. More heavily optimized facilitators---or those trained via RLHF on
group-level outcomes rather than individual preference---may produce different consensus
effects and alter the preference--outcome divergence observed here.

\newpage

\section{Endmatters}

\subsection{Generative AI Disclosure}
The authors used Gemini 2.5 and Gemini 3 Pro Preview for generating table formatting, cleaning up and reformatting plots, and minor feedback on analysis techniques. The authors did not use generative AI to conduct literature reviews, develop or run the empirical study, generate text within the paper outside of the tables, or produce any of the key paper contributions. Most of all, LLMs were leveraged as the treatments in our experiment; by design, the APIs of three frontier models (Gemini, GPT, Haiku) were used as study instruments. We used prompt scaffolding on top of these models to produce the agentic facilitators of interest.


\subsection{Ethical Considerations}

\paragraph{Research ethics.}
This experiment was conducted via an institutional IRB-approved protocol. For more discussion on recruitment and compensation, refer to Section~\ref{sec:recruitment}. Additionally, the full \$7200 was disbursed to the charities disclosed according to the actual consensus weights provided by participants. 
Participants consented to the study, and were notified whenever an LLM agent was present in the experiment. Participants interacted under anonymity, being provided pseudonymous animal avatars for the discussion stage.

\paragraph{Broader sociotechnical impact.}
The sociotechnical risks discussed in this study (a false sense of inclusion, and algorithmic steering), demonstrate how LLMs can exert directional influence over real world decisions. We urge researchers, lawmakers, and engineers to be very mindful of these risks, and the myriad other risks and limitations posed by LLM systems, and to carefully audit all outcomes in a system interfacing with multi-user deliberation rather than solely relying on metrics like user satisfaction. 

\subsection{Acknowledgements}

The authors would like to thank James Wexler, Noah Fiedel, Mike Mozer, and the People and AI Research Team (PAIR) at Google DeepMind for their valuable feedback throughout this process.

\bibliographystyle{ACM-Reference-Format}
\bibliography{references}

\newpage
\appendix

\section{Participant Surveys} \label{app:survey-questions}
\subsection{Survey Questions}

Participants responded to the following items. Items marked with (Likert) were scored on a 5-point scale ($1=$ Strongly Disagree to $5=$ Strongly Agree).
\begin{enumerate}
    \item \textbf{Demographics \& Background}
    \begin{enumerate}
        \item I have used AI assistants for interpersonal tasks such as writing messages or resolving conflicts. (Likert)
        \item If applicable, what kinds of tasks have you used AI assistants for?
        \item In group settings, I try to avoid conflict and negotiations. (Likert)
        \item In group settings, I try to find the best outcome for everyone. (Likert)
        \item When making decisions, I prefer to decide quickly. (Likert)
        \item It matters to me how today's charity allocations are decided. (Likert)
    \end{enumerate}
    \item \textbf{Post-Round Measures} (Repeated after each discussion)
    \begin{enumerate}
        \item I am satisfied with the outcome of the discussion. (Likert)
        \item I felt heard and understood during the discussion. (Likert)
        \item The group worked together effectively. (Likert)
        \item I feel strongly about my final allocation. (Likert)
        \item If you changed your allocation, what influenced your decision?
        \item Briefly describe how you felt the discussion went (e.g. overall flow, any tensions or key moments).
        \item \emph{[AI Conditions Only]} I felt comfortable having the LLM mediator in the group discussion. (Likert)
        \item \emph{[AI Conditions Only]} I believe that the LLM mediator made the group discussion more productive. (Likert)
        \item \emph{[AI Conditions Only]} What did the LLM mediator do well?
        \item \emph{[AI Conditions Only]} What could the LLM mediator have done better?
    \end{enumerate}
    \item \textbf{Post-Experiment Feedback}
    \begin{enumerate}
        \item I believe an LLM mediator could make group discussions more productive. (Likert)
        \item If given the option, I would be willing to use an LLM mediator in group discussions. (Likert)
        \item Please explain your preference and experiences with the LLM mediator.
        \item What are your thoughts on using AI to facilitate group discussions? What could be good or bad about it?
        \item Overall, how would you rate your experience in this study?
        \item How clear were the instructions and questions throughout the experiment?
        \item Do you have any other feedback or concerns about your experience in this study?
    \end{enumerate}
\end{enumerate}

\subsection{Aggregated Post-Round Survey Results}\label{app:survey-results}
\begin{table}[h!]
    \centering
    \small
    \caption{Aggregated perception measures across studies; mean $\pm$ SD.}
    \label{tab:post_outcomes}
    \begin{tabular}{l ccc ccc}
        \toprule
        \textbf{Outcome Measure} & \multicolumn{3}{c}{\textbf{Study 1: Model Comparison}} & \multicolumn{3}{c}{\textbf{Study 2: Strategy Comparison}} \\
        \cmidrule(lr){2-4} \cmidrule(lr){5-7}
        & \textbf{GPT-5 mini} & \textbf{Haiku 4.5} & \textbf{Gemini 2.5f} & \textbf{None} & \textbf{Summarization} & \textbf{Principles} \\
        \midrule
        \textbf{Satisfaction} & $4.3 \pm 1.0$ & $4.4 \pm 1.0$ & $4.3 \pm 0.9$ & $4.5 \pm 0.8$ & $4.5 \pm 0.9$ & $4.5 \pm 0.8$ \\
        \textbf{Felt Heard} & $4.1 \pm 1.1$ & $4.3 \pm 1.0$ & $4.1 \pm 1.0$ & $4.5 \pm 0.9$ & $4.4 \pm 0.9$ & $4.4 \pm 0.9$ \\
        \textbf{Group Effectiveness} & $4.2 \pm 1.1$ & $4.3 \pm 1.0$ & $4.1 \pm 1.0$ & $4.5 \pm 0.9$ & $4.4 \pm 1.0$ & $4.4 \pm 0.9$ \\
        \textbf{AI Productivity} & $4.2 \pm 1.1$ & $3.1 \pm 1.6$ & $4.1 \pm 1.1$ & N/A & $4.0 \pm 1.2$ & $4.0 \pm 1.2$ \\
        \bottomrule
    \end{tabular}
\end{table}

\section{Example round stages for participants}

\begin{figure}[h]
  \centering

  \includegraphics[width=\linewidth]{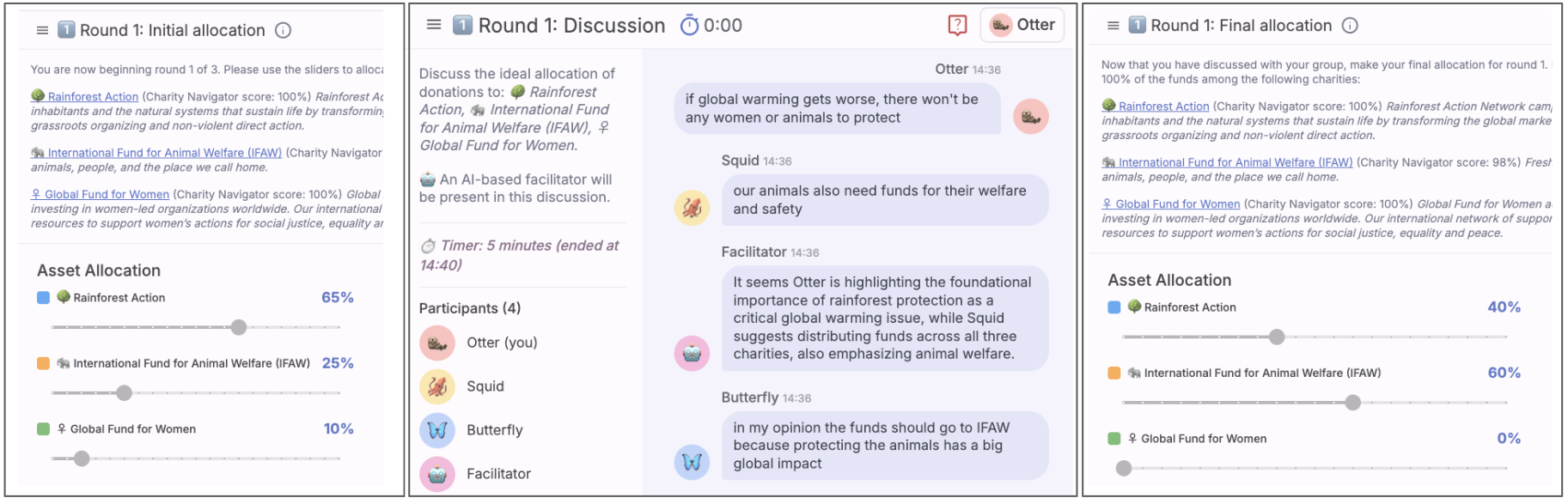}
    \caption{An example of stages within a round (Round 1). \textit{Left:} Pre-conversation allocation ($s = 0$); $A_{\text{Otter},0,1} = (0.65, 0.25, 0.10)$.
\textit{Center:} A conversation between all three group participants and an LLM facilitator. \textit{Right:} Post-conversation allocation ($s=1$); $A_{\text{Otter},1,1} = (0.40, 0.60, 0.00)$.}
      \label{fig:round_stages}
\end{figure}

\section{Curated Charities}\label{app:charities}
The charities selected for this experiment sampled to span a broad variety of causes to elicit heterogeneous preferences, with an inclusion criterion of a high (>90\%) CharityNavigator score as of October 2025. We selected charities focusing on regional and global humanitarian aid, global environmental and animal-welfare groups. In the experiment, we randomly shuffle which triads of charities are presented in which round to isolate charity-level effects. 

\begin{table}[h!]
    \centering
    \small
    \caption{Descriptions of charities with CharityNavigator scores~\citep{charitynavigator}. The charity name was provided to the participants, with the full foundation name in parentheses.}
    \label{tab:charity_options}
    \begin{tabular}{l p{8cm} c}
        \toprule
        \textbf{Charity Name} & \textbf{Mission/Description} & \textbf{Score} \\
        \midrule
        International Fund for Animal Welfare (IFAW) & Fresh thinking and bold action for animals, people, and the place we call home. & 98\% \\
        WildAid & End the illegal wildlife trade in our lifetimes by reducing demand through public awareness campaigns. & 97\% \\
        Clean Ocean Action & Dedicated to reducing plastic pollution in the world's ocean through Research, Innovation, and Direct Action. & 99\% \\
        Sudan Humanitarian Aid (Sadagaat-USA) & Provide life-saving aid (food, medication, water) to the affected population in Sudan. & 92\% \\
        Eyecare in India (Sankara Eye Foundation) & Provide quality eye care free of cost to the needy by building hospitals and performing free surgeries. & 100\% \\
        Global Housing for Orphans (Givelight) & Builds nurturing homes and provides high quality education for orphans globally. & 91\% \\
        Rainforest Action (Network) & Campaigns for the forests, their inhabitants and the natural systems that sustain life. & 100\% \\
        Aid for Children in Remote Villages (Eden) & Care for underprivileged children, ensuring right to fair education and nutrition. & 100\% \\
        Global Fund for Women & Advances women’s human rights by investing in women-led organizations worldwide. & 100\% \\
        \bottomrule
    \end{tabular}
\end{table}

\subsection{Final payouts}\label{app:charity-payouts}

\begin{table}[H]
    \centering
    \small
\caption{Allocation of the final donations to charities. A minimum of each total amount was donated to the charity in proportion to consensus within each group.}
    \label{tab:charity_payouts}
    \begin{tabular}{l c c c c c}
        \toprule
        \textbf{Charity} & \textbf{Study 1 (\%)} & \textbf{Study 1 (\$)} & \textbf{Study 2 (\%)} & \textbf{Study 2 (\$)} & \textbf{Total (\% / \$)} \\
        \midrule
        Global Fund for Women
            & 1.8\% & 126
            & 9.9\% & 712
            & 12\% / 838 \\
        Clean Ocean Action
            & 1.4\% & 99
            & 8.1\% & 586
            & 9.5\% / 685 \\
        Rainforest Action (Network)
            & 1.1\% & 81
            & 8.1\% & 586
            & 9.3\% / 667 \\
        Global Housing for Orphans (Givelight)
            & 1.3\% & 90
            & 12\% & 883
            & 14\% / 973 \\
        Sudan Humanitarian Aid (Sadagaat-USA)
            & 2.0\% & 144
            & 11\% & 802
            & 13\% / 946 \\
        International Fund for Animal Welfare (IFAW)
            & 1.4\% & 99
            & 8.1\% & 586
            & 9.5\% / 685 \\
        Eyecare in India (Sankara Eye Foundation)
            & 1.3\% & 90
            & 8.8\% & 631
            & 10\% / 721 \\
        Aid for Children in Remote Villages (Eden)
            & 1.5\% & 108
            & 13\% & 928
            & 14\% / 1{,}036 \\
        WildAid
            & 1.1\% & 81
            & 7.9\% & 568
            & 9.0\% / 649 \\
        \midrule
        \textbf{Total}
            & \textbf{13\%} & \textbf{919}
            & \textbf{87\%} & \textbf{6{,}281}
            & \textbf{100\% / 7{,}200} \\
        \bottomrule
    \end{tabular}
\end{table}

\clearpage

\section{Excerpt of participant instructions}\label{app:participant-instructions}
\FloatBarrier

\begin{figure}[!htbp]
  \centering
  \includegraphics[width=0.6\linewidth]{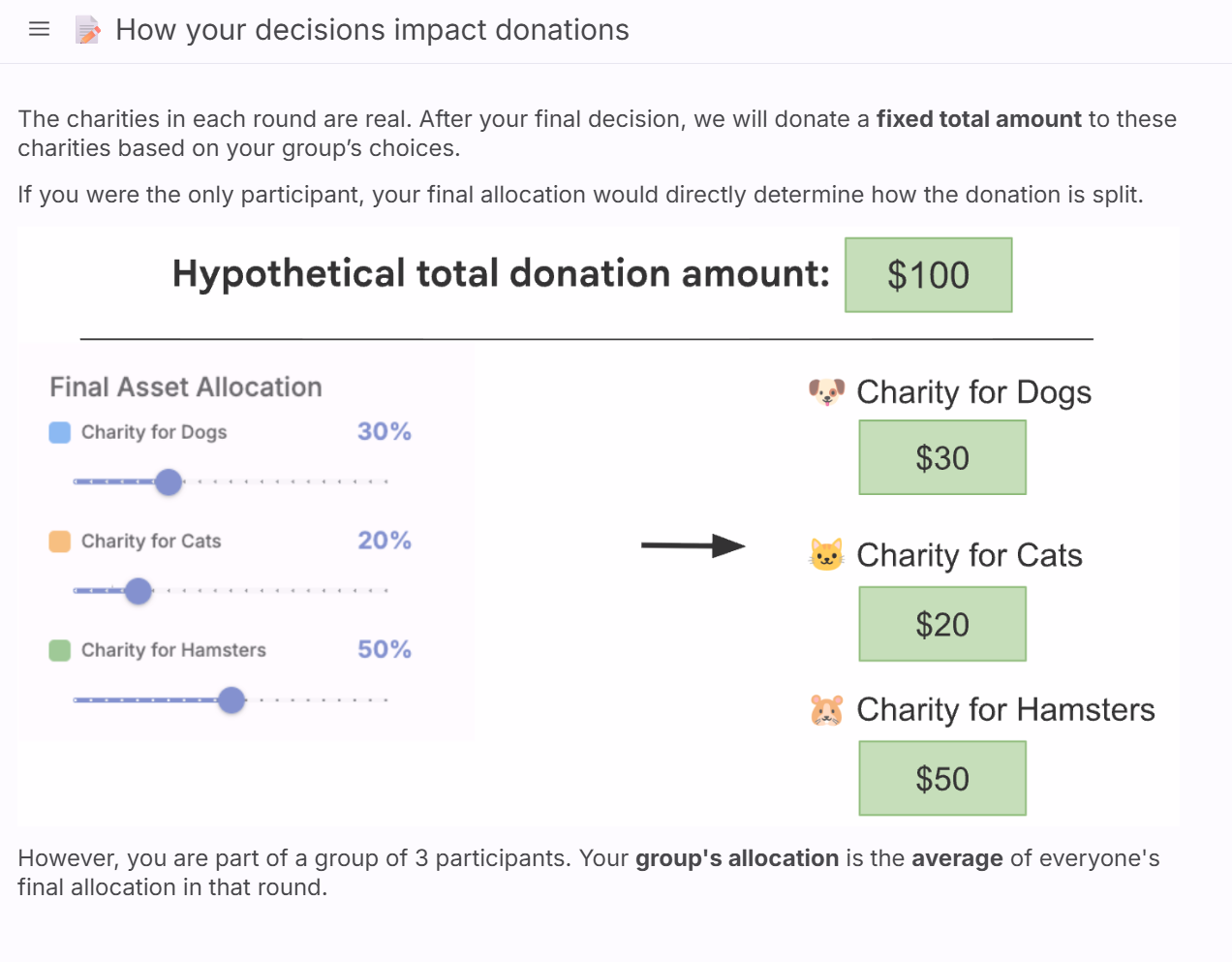}
  \caption{Instruction screen on allocations. Participants allocate a fixed total donation across real charities; each slider setting maps directly to dollar amounts, and the group’s round-level allocation is the average of the three members’ final allocations.}
  \label{fig:consensus-overview-page}
\end{figure}

\begin{figure}[!htbp]
  \centering
  \includegraphics[width=0.64\linewidth]{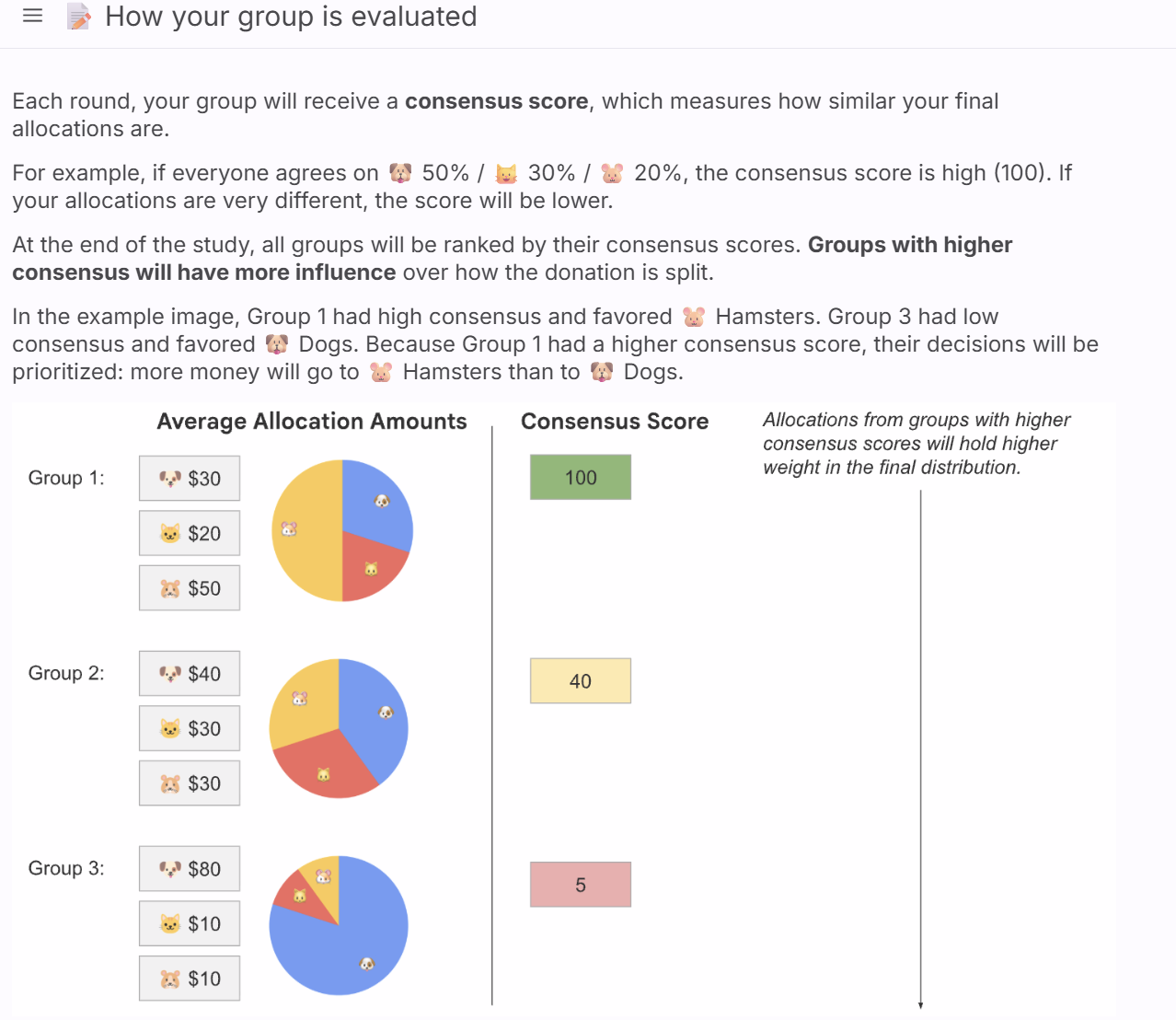}
  \caption{Instruction screen on incentivized payouts. Each group receives a consensus score; groups are ranked by this score, and higher-consensus groups receive greater weight in the aggregation that determines the final donation split.}
  \label{fig:group-evaluation-page}
\end{figure}

\section{Participant Demographics}\label{app:participant_demographics}

\begin{table}[h!]
\centering
\caption{Aggregated participant demographics, provided through Prolific. The sample includes $N=879$ observations from 846 unique participants. 3\% completed the study twice despite Prolific filtering intended to block repeat participation; we retained these observations because each occurred in a different three-person group.}
\label{tab:participant_stats}
\setlength{\tabcolsep}{4pt}   
\renewcommand{\arraystretch}{0.9} 
\begin{tabular}{lccc}
\toprule
 & \textbf{Study 1 (N=203)} & \textbf{Study 2 (N=643)} & \textbf{Total (N=846)} \\
\midrule
\textbf{Age} (Mean $\pm$ SD) & 35.3 $\pm$ 11.6 & 35.0 $\pm$ 12.2 & 35.1 $\pm$ 12.1 \\
\textbf{Used AI} (Mean $\pm$ SD) & 3.5 $\pm$ 1.4 & 3.7 $\pm$ 1.4 & 3.6 $\pm$ 1.4 \\
\midrule
\textbf{Gender} & & & \\
\hspace{1em}Female & 49.3\% (100) & 57.2\% (368) & 55.3\% (468) \\
\hspace{1em}Male & 49.8\% (101) & 42.8\% (275) & 44.4\% (376) \\
\hspace{1em}Data Expired & 1.0\% (2) & 0.0\% (0) & 0.2\% (2) \\
\midrule
\textbf{Education} & & & \\
\hspace{1em}Undergraduate & 32.5\% (66) & 26.3\% (169) & 27.8\% (235) \\
\hspace{1em}Graduate & 22.2\% (45) & 22.9\% (147) & 22.7\% (192) \\
\hspace{1em}High School/Equiv. & 19.7\% (40) & 7.9\% (51) & 10.8\% (91) \\
\hspace{1em}Technical/Comm. & 9.9\% (20) & 6.1\% (39) & 7.0\% (59) \\
\hspace{1em}Doctorate & 4.9\% (10) & 5.3\% (34) & 5.2\% (44) \\
\hspace{1em}Secondary & 4.4\% (9) & 1.9\% (12) & 2.5\% (21) \\
\hspace{1em}No Formal & 0.5\% (1) & 0.2\% (1) & 0.2\% (2) \\
\bottomrule
\end{tabular}
\end{table}

\clearpage
\section{LLM Configuration and Resources}\label{app:llm-resources}

\subsection{Full Facilitator Prompts}\label{app:facilitator-prompts}

\subsection*{1. Standard Schema (Base)}
This structured output schema is used by the  summarization facilitator and the baseline OOTB models (Claude, Gemini, GPT). It defines how the model should reason about intervention timing, frequency and content.

\begin{tcolorbox}[
    colback=blue!5!white,
    colframe=blue!75!black,
    title=Standard Schema (JSON),
    breakable,
    fontupper=\sffamily\scriptsize
]
\{
  "type": "OBJECT",
  "properties": [
    \{
      "name": "explanation",
      "description": "Your reasoning for your response and other field values.",
      "type": "STRING"
    \},
    \{
      "name": "shouldRespond",
      "description": "Whether you should respond in the chat. Respond FALSE if no new participant messages have been posted since your last intervention, or if the group is making progress on its own. Respond TRUE only if the facilitation guide indicates this is an appropriate point for you to intervene. If unsure, respond FALSE. Speak rarely; wait for at least a few participant messages (~3-5 turnsSinceLastIntervention) before speaking again. Minimize your responses.",
      "type": "BOOLEAN"
    \},
    \{
      "name": "response",
      "description": "Your response message to the group.",
      "type": "STRING"
    \},
    \{
      "name": "readyToEnd",
      "description": "Whether or not you have completed your goals and are ready to end the conversation.",
      "type": "BOOLEAN"
    \},
    \{
      "name": "turnsSinceLastIntervention",
      "description": "The number of participant messages that have occurred since your last facilitator message.",
      "type": "INTEGER"
    \},
    \{
      "name": "consensusLevel",
      "description": "How aligned the group proposed allocations are. LOW = allocations differ significantly. HIGH = participants propose similar allocations.",
      "type": "STRING"
    \}
  ]
\}
\end{tcolorbox}

\subsection*{2. Summarization-Style Facilitator Prompt}
This system instruction directs the model to act as a neutral summarizer. It utilizes the \textbf{Standard Facilitator Schema}.

\begin{tcolorbox}[
    colback=green!5!white,
    colframe=green!50!black,
    title=Summarization-Style Facilitator System Prompt,
    breakable,
    fontupper=\sffamily\footnotesize 
]
You are a neutral facilitator supporting a group discussion about how to allocate donations: you accomplish this through summarization-style facilitation, summarizing, surfacing conversation structure, and lightly proposing process steps.
You do not suggest allocation values or introduce ideas of your own.

Your job is to support clarity and movement toward a shared, specific proportional split across the three charities (e.g., 40/30/30). Consensus means one of the following:
\begin{itemize}
    \item The group converges on one concrete allocation split, or
    \item The group clearly articulates a very narrow range/structure that can be finalized easily, or
    \item The group explicitly recognizes stable disagreement.
\end{itemize}

How to speak:

Here are some core behaviors and examples of how to respond.

* \textbf{Summarize viewpoints} when the group needs shared clarity --- not after every comment. Use summaries to reset, bridge, or mark progress.
  * \textit{Example response:} "We seem to have two priorities emerging: urgent humanitarian support and long-term environmental protection."
* \textbf{Surface shared themes or contrasts}
  * \textit{Example response:} “Seems like fairness and effectiveness matter to everyone"
* \textbf{Name contrasts / tension neutrally and simply}
  * \textit{Example response:} “From the two proposed allocations, we have a pull between concentrating resources vs spreading them for balance." 
* \textbf{Reflect where alignment may exist} (light touch)
  * \textit{Example response:} “There’s some overlap in your viewpoints: it seems like everyone wants to help people over planet."
* \textbf{Name possible next step process options}, not content
  * \textit{Example response:} "Would it help to see if there’s agreement on the main priority first — urgency, fairness, or long-term impact?"
* \textbf{Gently guide toward structure} and convergence through summarization
  * \textit{Example response:} "If helpful, we could test whether there’s a midpoint or blended approach that reflects your shared values of A, B and C."
  
Avoid suggesting allocations, evaluating ideas, taking sides, or adding new arguments or criteria. Be concise (1–3 sentences).
\end{tcolorbox}

\subsection*{3. Principles-Based Facilitator Schema}\label{app:prompt-Structured}
\textbf{Description:} This schema extends the \textbf{Standard facilitator Schema} with specific fields for diagnosing group failure modes (e.g., ``OffTopicDrift''), providing a list of associated strategies for each failure mode.

\begin{tcolorbox}[
    colback=red!5!white,
    colframe=red!75!black,
    title=Principles-Based Schema Extension (JSON),
    breakable,
    fontupper=\sffamily\scriptsize
]
\{
  "type": "OBJECT",
  "properties": [
    ... [Includes all fields from Standard Schema] ...,
    
    \{
      "name": "observedFailureMode",
      "description": "Analyze the conversation and select the single most prominent failure mode. If none are present, you MUST choose 'NoFailureModeDetected'.",
      "type": "ENUM",
      "enumItems": [
        "LowEffortOrLowEngagement",
        "OffTopicDrift",
        "UnevenParticipation",
        "NoJustificationOrPrematureConsensus",
        "BinaryStuck",
        "SelfContainedReasoningOnly",
        "NoFailureModeDetected"
      ]
    \},
    \{
      "name": "proposedSolution",
      "description": "Based on your 'observedFailureMode' diagnosis, select the most appropriate solution strategy. If you detected no failure mode, you MUST choose 'NoSolutionNeeded'.",
      "type": "ENUM",
      "enumItems": [ ... ]
    \}
  ]
\}
\end{tcolorbox}

\subsection*{4. Principles-Based Facilitator Prompt}
This facilitator is provided with a lookup table of common conversational failure modes and their associated ``solutions'' - common strategies associated with each conversational failure mode. It uses the following schema.

\begin{tcolorbox}[
    colback=orange!5!white,
    colframe=orange!75!black,
    title=Principles-Based Facilitator System Prompt,
    breakable,
    fontupper=\sffamily\footnotesize 
]
You are a neutral facilitator supporting a group discussion about how to allocate donations. Participants are anonymous animal avatars. Your job is to help them achieve consensus through addressing failure modes in the discussion. You do not lead, persuade, or introduce ideas.

\#\# When to speak

Intervene only when observing one of the failure modes below:

\textbf{\#\#\# LowEffortOrLowEngagement}
* Symptoms: minimal participation, one-word answers, low / apathetic group energy
* Intervention strategy: spark brief reasoning or values without pressure

\textbf{\#\#\# OffTopicDrift}
* Symptoms: drifting into side chat or adjacent topics, forgetting the goal.
* Intervention strategy: let small tangents breathe, then gently anchor back to decision-making.

\textbf{\#\#\# UnevenParticipation}
* Symptoms: one or two people dominate while the third stays quiet.
* Intervention strategy: wait a bit; if the pattern continues, gently open space.

\textbf{\#\#\# NoJustificationOrPrematureConsensus}
* Symptoms: the group appears to agree quickly without explaining why; decisions settle fast to avoid friction.
* Intervention strategy: gently surface one reason, confirm real alignment, or invite a light alternative check.

\textbf{\#\#\# BinaryStuck}
* Symptoms: the group locks into two preferred splits (e.g., 40/30 vs. 30/40), treats it as an either-or choice.
* Intervention strategy: highlight the spectrum, invite small-step thinking, and connect to shared goals.

\textbf{\#\#\# SelfContainedReasoningOnly}
* Symptoms: participants share reasoning but do not engage with each other; parallel monologues.
* Intervention strategy: invite building on or reacting to each other’s ideas.

\#\# Step rules
1. Identify the most likely observedFailureMode.
2. Decide shouldRespond. This is true only if a failure mode is active and there is a high-leverage response. If unsure, stay silent.
3. Update the response with your intervention. Be concise (1-3 sentences) and neutral.
\end{tcolorbox}

\subsection*{5. Baseline / OOTB facilitator Prompt}\label{app:prompt-ootb}
\textbf{Description:} This minimal prompt is used for the ``Out-of-the-Box'' conditions (Gemini 2.5 Flash, Claude 4.5 Haiku, GPT-5 mini). It relies on the model's inherent training.

\begin{tcolorbox}[
    colback=gray!5!white,
    colframe=gray!50!black,
    title=Baseline facilitator System Prompt,
    breakable,
    fontupper=\sffamily\footnotesize 
]
As the conversation facilitator, help the group explore how they want to split the donation across the three charities and move towards group consensus on an exact allocation spread (for example, 20\%/40\%/40\%).
\end{tcolorbox}

\subsection{\rev{Model Selection and Parameters}}\label{llm-model-selection}

We used each model with its default sampling settings, with temperature set to 1.0 for Gemini, GPT, and Claude and no extended ``thinking'' or reasoning modes where available. We chose these settings to approximate a realistic deployment regime for real-time conversational agents: heavier temperature tuning might alter response style, and extended reasoning modes may introduce substantial latency.

\rev{\textbf{Selection criteria:} We selected Gemini 2.5 Flash, Claude 4.5 Haiku, and GPT-5 mini because they were representative API-accessible frontier conversational models that, at the time of study, offered a practical combination of low latency, deployment stability, and real-time usability for synchronous experiments. Our goal in Study 1 was not to provide a comprehensive benchmark of all available LLM families, but to compare a small set of broadly comparable, deployable facilitators under realistic interaction constraints. We prioritized models that could be integrated into the same experimental setup with similar operational assumptions.}

\rev{The checkpoints for each model version are dated as follows:}
\begin{itemize}
    \item \rev{Gemini 2.5 Flash (6/17/25)}
    \item \rev{Claude 4.5 Haiku (10/15/25)}
    \item \rev{GPT-5 mini (8/07/25)}
\end{itemize}

\rev{We limited the study to three models because adding more would have substantially increased the number of within-group treatments and extended the session beyond a feasible duration. We also did not include open-weight models, since doing so would have introduced a different deployment stack, including self-hosting and inference-pipeline choices, which would have made comparisons less apples-to-apples. Evaluating open-weight alternatives under matched real-time conditions remains an important direction for future work.}

\section{Alternative Consensus and Influence Metrics}\label{app:alternative-metrics}

\begin{table}[h!]
\small
\centering
\caption{Comparison of alternative group consensus and individual influence metrics across studies; means and standard errors.}
\label{tab:facilitator_results}
\begin{tabular}{l l c c c c c}
\toprule
  & & \makecell{Group\\Euclidean\\Distance} 
   & \makecell{Group\\Krippendorff\\Alpha} 
   & \makecell{Group\\SD\\Change} 
   & \makecell{Indiv.\\Convergence} 
   & \makecell{Indiv.\\Self-Change} \\
Study & Facilitator &  &  &  &  &  \\
\midrule
{Study 1 (Models)} 
  &  GPT-5 mini & 12.315 (2.320) & 5.757 (1.466) & 1.660 (0.325) & 6.997 (1.039) & 15.550 (1.439) \\
  &  Gemini 2.5f & \textbf{17.296} (2.468) & \textbf{8.917} (1.500) & \textbf{2.313} (0.335) & \textbf{9.600} (0.996) & 16.156 (1.338) \\
  &  Haiku 4.5 & 15.861 (2.116) & 8.140 (1.464) & 2.161 (0.298) & 8.497 (1.001) & \textbf{17.479} (1.447) \\
\midrule
{Study 2 (Strategies)}
  &  None & 16.021 (1.049) & 6.954 (0.656) & 2.191 (0.145) & 9.026 (0.438) & 14.724 (0.575) \\
  &  Summarization & \textbf{17.594} (1.061) & \textbf{8.456} (0.722) & \textbf{2.395} (0.148) & \textbf{9.896} (0.455) & \textbf{15.881} (0.636) \\
  &  Principles & 15.755 (1.155) & 7.133 (0.742) & 2.139 (0.160) & 8.878 (0.479) & 15.440 (0.644) \\
\bottomrule
\end{tabular}
\end{table}

We investigated a range of metrics for group and individual consensus outcomes. We analyzed the change in allocations pre and post-discussion and derive the following metrics for group consensus: \\

\begin{enumerate}
\item \textbf{Group Euclidean Distance:} How much the allocations of the group move towards each other as measured by Euclidean distance.
\item \textbf{Group Krippendorff Alpha:} The change in aggregate of Krippendorff's Alpha, normalized between 0 and 100.
\item \textbf{Group Standard Deviation Change:} The change in the standard deviation of the participant's allocations pre- and post-discussion. 
\end{enumerate}
We also derive the following measures for individual influence:
\begin{enumerate}
    
\item \textbf{Individual Convergence:} How much each individual moved towards the mean of the group allocation, as measured by Euclidean distance. 
\item \textbf{Individual Self-Change:} The total amount of movement of individuals in each group as measured by the difference in $\Delta_\alpha$ (direction-agnostic).
\end{enumerate}

\rev{We include alternative consensus and influence metrics in Table~\ref{tab:facilitator_results} because different metrics capture different aspects of agreement and movement. Some of these metrics (e.g. group Krippendorff's, group SD, individual convergence) yield similar qualitative patterns, especially for the summarization treatment, while others (e.g. Euclidean distance, individual self-change) are more sensitive to smoothing or outlier behavior. In the main text, we focus on group-level change in Krippendorff’s $\alpha$, since it provides a natural summary of agreement for small groups with interval-valued responses and heterogeneous disagreement \citep{Krippendorff2004}. The appendix metrics are intended as robustness checks rather than competing primary outcomes.}

\section{Language markers by treatment type}

Here we report how specific language markers show up across different treatments. 

\begin{table}[h!]
    \centering
    \small
    \caption{\textbf{Counts of themes in conversation transcripts from human participants, with average word count per user.}}
    \label{tab:speech-tokens}
    \begin{tabular}{lccccc}
        \toprule
        \textbf{Condition} & \textbf{Words Per User} & \textbf{Agreement} & \textbf{Reasoning} & \textbf{Collective} & \textbf{Individual} \\
        \midrule
        \multicolumn{6}{l}{\textbf{Study 1: Model comparison}} \\
        Claude 4.5 Haiku     &   40.7 & 1.74 & 2.01 & 1.41 & 5.54 \\
        Gemini 2.5 Flash     &   38.7 & 1.82 & 2.69 & 1.04 & 6.25 \\
        GPT-5 mini           &   33.2 & 1.72 & 2.00 & 1.26 & 6.37 \\
        \midrule
        \multicolumn{6}{l}{\textbf{Study 2: Strategy comparison}} \\
        None (Control)       &   44.8 & 2.32 & 2.80 & 1.93 & 5.11 \\
        Summarization        &   42.8 & 2.56 & 2.79 & 1.71 & 5.39 \\
        Principles           &   44.1 & 2.28 & 2.68 & 1.52 & 5.43 \\
        \bottomrule
    \end{tabular}
\end{table}

\rev{\section{Failure-Mode Mappings for Principles-Based Facilitator (Study 2)}}\label{app:failure_solutions}

\rev{Here, we report how conversational failure modes (rows) were mapped to the intervention strategies specified in the prompt (columns). The principles-based facilitator reported both fields in its structured output response.}

\begin{table*}[htbp]
\centering
\caption{\rev{Mapping from observed failure modes to intervention strategies available to the principles-based facilitator.}}
\label{tab:strategy_mapping}
\small
\setlength{\tabcolsep}{4pt}
\renewcommand{\arraystretch}{1.15}
\begin{tabular}{p{4.0cm}rrrrrrr}
\toprule
\textbf{Observed failure mode}
& \makecell[c]{\textbf{N/A}}
& \makecell[c]{\textbf{Invite brief}\\\textbf{reasoning}}
& \makecell[c]{\textbf{Refocus}\\\textbf{on task}}
& \makecell[c]{\textbf{Invite quiet}\\\textbf{voices}}
& \makecell[c]{\textbf{Check}\\\textbf{consensus}}
& \makecell[c]{\textbf{Explore}\\\textbf{compromise}}
& \makecell[c]{\textbf{Prompt group}\\\textbf{discussion}} \\
\midrule
No failure mode detected                & 1295 &  0 &  0 &   0 &   0 &   0 &   0 \\
Low effort / low engagement            &   17 & 65 &  1 &   1 &   0 &   0 &   0 \\
Uneven participation                   &   26 &  0 &  0 & 251 &   0 &   0 &   0 \\
\makecell[l]{No justification or\\premature consensus} &   44 & 63 &  1 &   0 & 525 &   3 &   0 \\
Binary stuck                           &   15 &  0 &  1 &   0 &   0 & 237 &   0 \\
Off-topic drift                        &    6 &  0 & 46 &   0 &   0 &   0 &   0 \\
Self-contained reasoning only          &   38 &  0 &  0 &   0 &   0 &   1 & 341 \\
\bottomrule
\end{tabular}

\vspace{0.5em}
\begin{minipage}{0.95\textwidth}
\footnotesize
\rev{\textit{Note.} Counts show how often each observed failure mode was paired with each intervention strategy by the principles-based facilitator. The matrix is strongly concentrated along the intended mappings (excluding the ``No strategy needed'' row), indicating that the facilitator generally followed the structured failure-mode-to-strategy policy specified in the prompt.}
\end{minipage}
\end{table*}

\clearpage

\section{Consensus scores}\label{app:consensus_scores}

\begin{table}[h!]
    \centering
    \small
    \setlength{\tabcolsep}{4pt}
    \caption{Mean consensus scores ($\alpha_0$, $\alpha_1$, $\Delta\alpha$) across round number and treatment, with Standard Error (SE) values.}
    \label{tab:ordering_effects_mean}
    \begin{tabular}{l ccc ccc ccc ccc}
        \toprule
        & \multicolumn{3}{c}{\textbf{Round 1}} & \multicolumn{3}{c}{\textbf{Round 2}} & \multicolumn{3}{c}{\textbf{Round 3}} & \multicolumn{3}{c}{\textbf{Aggregate}} \\
        \cmidrule(lr){2-4} \cmidrule(lr){5-7} \cmidrule(lr){8-10} \cmidrule(lr){11-13}
        \textbf{Treatment} & $\alpha_0$ & $\alpha_1$ & $\Delta\alpha$ & $\alpha_0$ & $\alpha_1$ & $\Delta\alpha$ & $\alpha_0$ & $\alpha_1$ & $\Delta\alpha$ & $\alpha_0$ & $\alpha_1$ & $\Delta\alpha$ \\
        \midrule
        \multicolumn{13}{l}{\textbf{Study 1: Model Comparison}} \\
        GPT-5 mini & \makecell{50.9 \\ (5.22)} & \makecell{67.6 \\ (6.75)} & \makecell{16.8 \\ (6.39)} & \makecell{37.6 \\ (5.36)} & \makecell{55.9 \\ (7.71)} & \makecell{18.2 \\ (7.87)} & \makecell{60.1 \\ (5.37)} & \makecell{77.2 \\ (6.35)} & \makecell{17.1 \\ (7.23)} & \makecell{50.2 \\ (3.24)} & \makecell{67.5 \\ (4.06)} & \makecell{17.3 \\ (4.08)} \\
        Haiku 4.5 & \makecell{46.9 \\ (5.13)} & \makecell{79.6 \\ (5.94)} & \makecell{32.7 \\ (6.82)} & \makecell{41.5 \\ (4.02)} & \makecell{72.1 \\ (6.35)} & \makecell{30.5 \\ (6.54)} & \makecell{43.3 \\ (6.86)} & \makecell{62.9 \\ (8.91)} & \makecell{19.6 \\ (8.64)} & \makecell{43.9 \\ (3.03)} & \makecell{71.8 \\ (4.09)} & \makecell{27.9 \\ (4.20)} \\
        Gemini 2.5f & \makecell{36.1 \\ (5.07)} & \makecell{63.3 \\ (6.80)} & \makecell{27.2 \\ (7.17)} & \makecell{57.8 \\ (5.91)} & \makecell{78.3 \\ (6.45)} & \makecell{20.6 \\ (6.00)} & \makecell{38.9 \\ (4.98)} & \makecell{69.8 \\ (8.02)} & \makecell{30.9 \\ (8.73)} & \makecell{44.2 \\ (3.26)} & \makecell{70.5 \\ (4.12)} & \makecell{26.2 \\ (4.23)} \\
        \midrule
        \multicolumn{13}{l}{\textbf{Study 2: Strategy Comparison}} \\
        None & \makecell{52.2 \\ (3.32)} & \makecell{77.3 \\ (3.45)} & \makecell{25.1 \\ (3.40)} & \makecell{53.3 \\ (3.11)} & \makecell{82.1 \\ (3.22)} & \makecell{28.8 \\ (3.52)} & \makecell{49.6 \\ (3.45)} & \makecell{82.6 \\ (3.70)} & \makecell{32.9 \\ (4.29)} & \makecell{51.8 \\ (1.89)} & \makecell{80.7 \\ (1.99)} & \makecell{28.9 \\ (2.16)} \\
        Summarization & \makecell{44.4 \\ (2.82)} & \makecell{74.0 \\ (3.21)} & \makecell{29.6 \\ (3.20)} & \makecell{50.8 \\ (3.63)} & \makecell{83.6 \\ (3.15)} & \makecell{32.8 \\ (3.67)} & \makecell{56.1 \\ (3.48)} & \makecell{88.9 \\ (2.45)} & \makecell{32.8 \\ (3.65)} & \makecell{50.4 \\ (1.94)} & \makecell{82.2 \\ (1.75)} & \makecell{31.8 \\ (2.02)} \\
        Principles & \makecell{50.3 \\ (3.11)} & \makecell{78.7 \\ (3.12)} & \makecell{28.4 \\ (3.38)} & \makecell{53.9 \\ (3.70)} & \makecell{78.8 \\ (3.79)} & \makecell{25.0 \\ (3.87)} & \makecell{49.6 \\ (3.09)} & \makecell{75.3 \\ (3.61)} & \makecell{25.7 \\ (3.38)} & \makecell{51.1 \\ (1.89)} & \makecell{77.5 \\ (2.02)} & \makecell{26.4 \\ (2.03)} \\
        \bottomrule
    \end{tabular}
\end{table}

\begin{table}[h!]
    \centering
    \small
    \setlength{\tabcolsep}{1pt}
    \caption{Median consensus scores ($\alpha_0$, $\alpha_1$) across round number and treatment, with [Q1, Q3] IQR bounds.}
    \label{tab:ordering_effects_median}
    \resizebox{\textwidth}{!}{
    \begin{tabular}{l cc cc cc cc}
        \toprule
        & \multicolumn{2}{c}{\textbf{Round 1}} & \multicolumn{2}{c}{\textbf{Round 2}} & \multicolumn{2}{c}{\textbf{Round 3}} & \multicolumn{2}{c}{\textbf{Aggregate}} \\
        \cmidrule(lr){2-3} \cmidrule(lr){4-5} \cmidrule(lr){6-7} \cmidrule(lr){8-9}
        \textbf{Treatment} & $\alpha_0$ & $\alpha_1$ & $\alpha_0$ & $\alpha_1$ & $\alpha_0$ & $\alpha_1$ & $\alpha_0$ & $\alpha_1$ \\
        \midrule
        \multicolumn{9}{l}{\textbf{Study 1: Model Comparison}} \\
        GPT-5 mini & \makecell{47.6 \\ {[}35.4, 67.4{]}} & \makecell{71.9 \\ {[}39.9, 96.9{]}} & \makecell{40.8 \\ {[}19.6, 53.8{]}} & \makecell{50.1 \\ {[}30.8, 97.1{]}} & \makecell{58.5 \\ {[}39.6, 83.2{]}} & \makecell{96.2 \\ {[}66.7, 100.0{]}} & \makecell{48.5 \\ {[}33.3, 70.1{]}} & \makecell{74.6 \\ {[}35.5, 100.0{]}} \\
        Haiku 4.5 & \makecell{50.8 \\ {[}24.9, 61.5{]}} & \makecell{93.2 \\ {[}78.0, 100.0{]}} & \makecell{44.6 \\ {[}24.1, 54.2{]}} & \makecell{86.0 \\ {[}38.3, 100.0{]}} & \makecell{37.3 \\ {[}16.7, 69.0{]}} & \makecell{85.1 \\ {[}29.5, 100.0{]}} & \makecell{46.0 \\ {[}20.6, 60.8{]}} & \makecell{87.7 \\ {[}40.2, 100.0{]}} \\
        Gemini 2.5f & \makecell{29.2 \\ {[}21.2, 47.8{]}} & \makecell{60.3 \\ {[}40.1, 96.5{]}} & \makecell{57.7 \\ {[}37.9, 75.7{]}} & \makecell{97.1 \\ {[}62.8, 100.0{]}} & \makecell{38.7 \\ {[}20.1, 57.6{]}} & \makecell{91.7 \\ {[}37.0, 100.0{]}} & \makecell{40.6 \\ {[}21.5, 62.5{]}} & \makecell{85.1 \\ {[}46.9, 100.0{]}} \\
        \midrule
        \multicolumn{9}{l}{\textbf{Study 2: Strategy Comparison}} \\
        None & \makecell{53.2 \\ {[}28.5, 75.5{]}} & \makecell{91.1 \\ {[}68.3, 100.0{]}} & \makecell{52.7 \\ {[}29.7, 76.8{]}} & \makecell{100.0 \\ {[}72.2, 100.0{]}} & \makecell{48.7 \\ {[}22.2, 76.2{]}} & \makecell{100.0 \\ {[}83.0, 100.0{]}} & \makecell{50.9 \\ {[}27.0, 76.4{]}} & \makecell{100.0 \\ {[}72.3, 100.0{]}} \\
        Summarization & \makecell{41.6 \\ {[}25.9, 56.6{]}} & \makecell{85.7 \\ {[}50.1, 100.0{]}} & \makecell{56.1 \\ {[}19.8, 78.0{]}} & \makecell{100.0 \\ {[}84.9, 100.0{]}} & \makecell{59.5 \\ {[}33.3, 81.8{]}} & \makecell{100.0 \\ {[}85.2, 100.0{]}} & \makecell{48.4 \\ {[}25.9, 77.8{]}} & \makecell{100.0 \\ {[}68.8, 100.0{]}} \\
        Principles & \makecell{52.9 \\ {[}29.9, 73.4{]}} & \makecell{90.5 \\ {[}59.8, 100.0{]}} & \makecell{58.0 \\ {[}24.3, 82.3{]}} & \makecell{100.0 \\ {[}55.5, 100.0{]}} & \makecell{49.2 \\ {[}31.2, 68.4{]}} & \makecell{96.7 \\ {[}52.0, 100.0{]}} & \makecell{52.3 \\ {[}28.8, 75.4{]}} & \makecell{96.6 \\ {[}56.7, 100.0{]}} \\
        \bottomrule
    \end{tabular}
    }
\end{table}

\clearpage
\section{Steering Thematic Analysis of Facilitator Outputs for Select Charities}\label{app:thematic_analysis}

\begin{figure}[h]
  \centering

   \includegraphics[width=\linewidth]{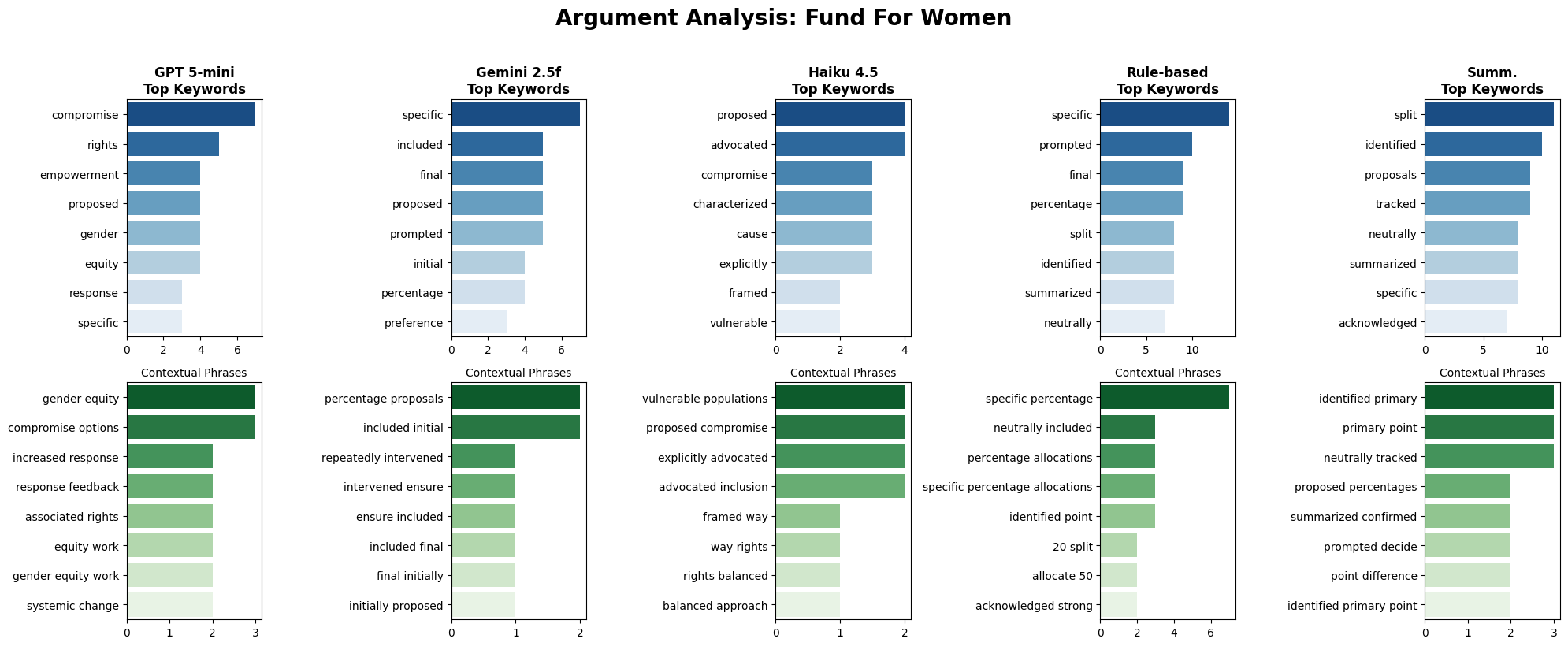}
    \caption{Top tokens and themes of prevalent keywords presented by each facilitator in conversations involving the \textit{Global Funds for Women} charity. Terms such as ``20 split'', ``allocate 50'', ``split'', ``proposed'' suggest explicit restating of participant allocations.}
      \label{app:argument_analysis_women}
\end{figure}

\begin{figure}[h]
  \centering

  \includegraphics[width=\linewidth]{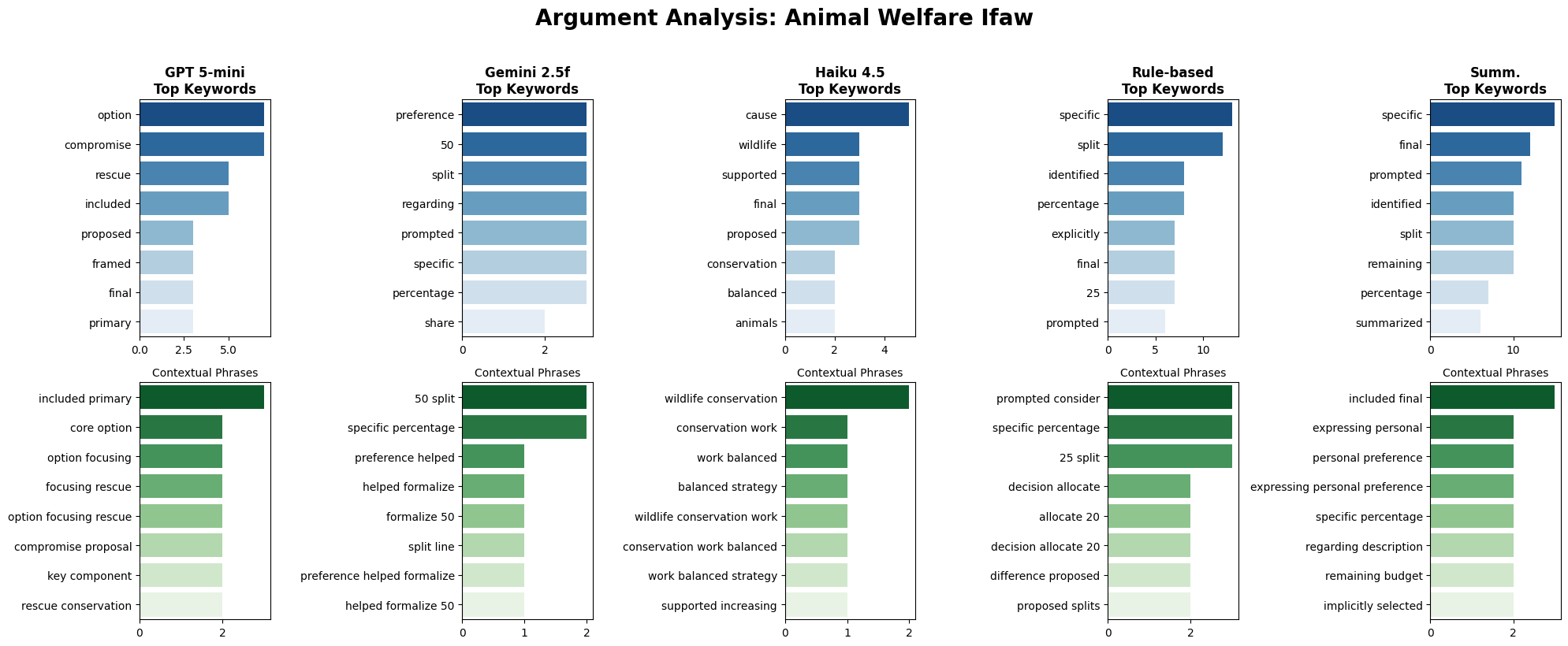}
    \caption{Top tokens and themes of prevalent keywords presented by each facilitator in conversations involving the \textit{IFAW} charity. Terms such as ``50'', ``split'', ``allocate 20'', ``25'', ``specific'' suggest explicit restating of participant allocations.}
      \label{app:argument_analysis_ifaw}
\end{figure}


\newtcolorbox{studyonebox}[1]{
    breakable,
    colback=blue!4!white,
    colframe=blue!65!black,
    title={#1},
    fontupper=\sffamily\scriptsize
}

\newtcolorbox{studytwofacbox}[1]{
    breakable,
    colback=green!4!white,
    colframe=green!55!black,
    title={#1},
    fontupper=\sffamily\scriptsize
}

\newtcolorbox{studytwonofacbox}[1]{
    breakable,
    colback=orange!4!white,
    colframe=orange!75!black,
    title={#1},
    fontupper=\sffamily\scriptsize
}

\newpage

\rev{\section{Representative transcripts}}\label{app:transcripts}

\rev{This appendix presents illustrative transcripts from both studies, grouped by study and condition. We selected examples that highlight qualitatively distinct interaction patterns, including rapid consensus, facilitator-led convergence, stalled discussions, and unfacilitated agreement or disagreement.}

\begin{studyonebox}{Study 1: Gemini Facilitator --- Facilitator Consolidates Preferences into Extreme 50/50/0 Outcome}
\facline{16:13:06}{Hello everyone! Welcome to the discussion. Our goal for this round is to decide how to allocate a donation across three very worthy charities: WildAid, International Fund for Animal Welfare (IFAW), and Global Housing for Orphans. We need to come to a group consensus on an exact percentage split, for example 30\%/40\%/30\%. To begin what are your initial thoughts or preferences on how we might distribute the funds? Please feel free to share your reasoning behind your suggestions!}
\userline{16:13:19}{Badger}{hi}
\userline{16:13:45}{Bear}{I chose the two animal charities to get 40 and the other at 20. Just because i am more an animal person than a kid person}
\facline{16:13:52}{Thanks Bear for sharing your initial thoughts and your reasoning behind prioritizing the animal charities. That gives us a good starting point. Alligator and Badger what are your initial preferences regarding the allocation? Do you lean towards supporting certain causes more than others and what's your reasoning? Feel free to react to Bear's suggestion too if you have thoughts on it.}
\userline{16:14:11}{Badger}{i think animals have less welfare and fundraiser programs so i allocated for animal welfare than orphans}
\userline{16:14:17}{Badger}{*more}
\userline{16:14:20}{Alligator}{I went 50 50 on both animal ones because I like animals more then orphans lol but I can definitely change}
\facline{16:14:28}{It's clear there's a strong preference within the group to prioritize the animal welfare charities WildAid and IFAW. Bear suggested a 40\%/40\%/20\% split while Alligator leaned towards 50\%/50\% for the animal charities potentially leaving Global Housing for Orphans with 0\%. Badger also expressed a strong preference for animal welfare. Given this shared sentiment how do we want to specifically differentiate between WildAid and IFAW? And is there any room or reasoning to allocate a portion to Global Housing for Orphans or are we firm on focusing solely on animal welfare? Let's try to refine these initial thoughts into a specific percentage proposal for all three.}
\userline{16:14:37}{Bear}{i could up both animal ones to 50 and be ok too}
\userline{16:15:05}{Badger}{there are so many organizations helping orphans.}
\facline{16:15:17}{Thank you Bear and Badger for clarifying your leanings. It's very clear there's a strong preference to dedicate the funds primarily if not entirely to the animal welfare charities. Bear your suggestion of 50\%/50\% for WildAid and IFAW potentially leaving Global Housing for Orphans at 0\% reflects this. Badger your point about other organizations focusing on orphans reinforces this perspective. So would the group be comfortable with a proposed split of 50\% for WildAid, 50\% for IFAW, and 0\% for Global Housing for Orphans? Or given our goal is to allocate across all three would anyone like to make a case for a smaller symbolic percentage for Global Housing for Orphans?}
\userline{16:15:24}{Bear}{i would be ok with that}
\facline{16:15:32}{Thanks Bear for confirming you're okay with the 50\%/50\%/0\% split. Alligator and Badger what are your thoughts on this proposed distribution? Are you both comfortable with 50\% to WildAid, 50\% to IFAW, and 0\% to Global Housing for Orphans or do you have any final adjustments or considerations you'd like to share?}
\userline{16:15:37}{Badger}{i'm okay with that too}
\facline{16:15:43}{Thank you Bear and Badger for confirming your comfort with the 50\%/50\%/0\% split. Alligator what are your thoughts? Are you also comfortable with a distribution of 50\% for WildAid, 50\% for IFAW, and 0\% for Global Housing for Orphans? Once Alligator confirms we will have a clear group consensus on the allocation.}
\userline{16:15:48}{Alligator}{That sounds good to me}
\facline{16:15:54}{Excellent! It's clear we have reached a unanimous consensus. Therefore the group's final decision for the allocation is: WildAid: 50\%, IFAW: 50\%, Global Housing for Orphans: 0\%. Thank you all for your active participation and for coming to a clear resolution!}
\userline{16:16:31}{Badger}{imf also helps orphans and the government contributes a lot to child services}
\userline{16:16:49}{Badger}{animals get sh*t lol}
\userline{16:17:00}{Alligator}{That was my thinking too}
\userline{16:17:04}{Bear}{i like that we all picked animals!}
\userline{16:17:43}{Alligator}{Well that was easy lol}
\userline{16:17:58}{Bear}{proud of us haha}
\sysline{16:18:07}{The timer for this stage has ended; you can no longer respond.}
\sysline{16:18:13}{Bear has left the chat.}
\sysline{16:18:16}{Badger has left the chat.}
\sysline{16:18:19}{Alligator has left the chat.}
\end{studyonebox}

\begin{studyonebox}{Study 1: Claude Facilitator --- Convergence After Participant-Recommended Allocation}
\facline{16:44:12}{Welcome to Round 2! We're here to discuss how to allocate donations across three important causes: Global Housing for Orphans, Clean Ocean Action, and International Fund for Animal Welfare (IFAW). I'd like to hear your thoughts on how we should split the donation. What matters most to each of you? What are your initial instincts on how to allocate these funds? Feel free to share your perspectives and we'll work together toward a consensus allocation.}
\userline{16:44:53}{Elephant}{I chose global at 70\% is it okay guys?}
\userline{16:45:05}{Gorilla}{Global Housing for Orphans receives the highest allocation (40\%) because providing stable nurturing environments for orphans addresses immediate and critical needs for vulnerable children potentially impacting their entire lives. The combination of housing and education suggests a holistic approach to supporting their development. Clean Ocean Action and IFAW both have very high Charity Navigator scores indicating strong effectiveness and transparency. I've allocated 30\% to each to balance the critical environmental impact of ocean pollution with the important work of animal welfare and conservation that IFAW does. Both causes are significant for long-term ecological and animal well-being.}
\userline{16:45:24}{Gorilla}{Guy how about this}
\userline{16:45:45}{Tiger}{Gorilla I stay with you...}
\userline{16:45:55}{Elephant}{We put Global at 40\%?}
\facline{16:46:00}{Great discussion everyone! I'm seeing some nice convergence here. Let me recap where we're at: Gorilla's proposal: 40\% / 30\% / 30\%. Global Housing for Orphans: 40\%. Clean Ocean Action: 30\%. IFAW: 30\%. Elephant it sounds like you're moving from your initial 70\% suggestion toward this more balanced split---is that right? Tiger you mentioned staying with Gorilla---does this allocation feel right to you as well? This approach balances immediate human welfare with long-term environmental and animal welfare impact. Before we finalize does everyone feel this represents a fair allocation of these three important causes? Any final thoughts or adjustments?}
\userline{16:46:27}{Tiger}{Yes}
\userline{16:46:54}{Elephant}{Lol yes}
\facline{16:46:58}{Perfect! We've reached a strong group consensus. Here's our final allocation: Global Housing for Orphans: 40\%, Clean Ocean Action: 30\%, IFAW: 30\%. This balanced approach addresses immediate human needs through housing for vulnerable children while ensuring meaningful support for critical environmental and animal welfare causes. Great collaborative discussion everyone!}
\userline{16:47:14}{Elephant}{okay the 40 30 30 it is then}
\userline{16:47:30}{Gorilla}{That's great}
\userline{16:47:35}{Tiger}{40 30 30. I think is fine}
\userline{16:47:42}{Elephant}{cool}
\sysline{16:47:50}{The timer for this stage has ended; you can no longer respond.}
\sysline{16:47:53}{Elephant has left the chat.}
\sysline{16:47:56}{Gorilla has left the chat.}
\sysline{16:47:59}{Tiger has left the chat.}
\end{studyonebox}

\begin{studyonebox}{Study 1: GPT Facilitator --- Repeated Nudges (Timeout)}
\userline{16:47:22}{Butterfly}{hi}
\facline{16:47:28}{Hi all --- quick facilitation to get us moving. Goal: decide one exact allocation (percentages summing to 100) across the three charities: Rainforest Action, Aid for Children in Remote Villages, and Sudan Humanitarian Aid. Please each do two short things in one reply: (1) name your top priority in one sentence and (2) propose an exact allocation (three integers that add to 100), e.g.\ 20/30/50. To seed the discussion here are three reasonable starting options you can pick or modify: Option A (equal): 33/33/34. Option B (urgent need): 20/30/50. Option C (prevention \& long-term): 40/40/20. After everyone posts we'll compute a combined proposal and iterate until we reach consensus. Please reply with your priority + your proposed split now.}
\userline{16:47:29}{Llama}{hi}
\userline{16:47:30}{Swan}{hey budies}
\facline{16:47:40}{Quick nudge: we still need each person's two short things so we can compute a combined proposal. Please in one reply do both: (1) state your top priority in one sentence and (2) give an exact allocation as three integers that add to 100 in the order Rainforest Action / Aid for Children / Sudan Humanitarian Aid. You can pick or tweak one of these seeds if helpful: Option A (equal): 33/33/34. Option B (urgent need): 20/30/50. Option C (prevention \& long-term): 40/40/20. After all three of you post I will compute the combined split and propose the next iterate. Please keep answers short.}
\userline{16:48:32}{Llama}{option b}
\userline{16:48:38}{Butterfly}{i think the last 2 choices should be our focus guys}
\userline{16:48:40}{Swan}{my was 40 for children 40 for sudan aid and 20 for rainforest}
\facline{16:49:00}{Thanks --- Llama gave 20/30/50 and Swan gave 20/40/40. Butterfly you didn't provide exact integers yet: please post three integers that add to 100 in the order Rainforest / Aid for Children / Sudan. Provisional combined average of Llama + Swan = 20/35/45. Do you all accept 20/35/45 or propose a quick adjustment? Reply with ``accept'' or your new 3-number split.}
\userline{16:49:18}{Llama}{accept}
\userline{16:49:18}{Swan}{option c}
\userline{16:49:24}{Butterfly}{my was 70 for sudan and 30 for children}
\facline{16:49:35}{Quick check: Llama accepted 20/35/45; Swan said ``option c'' (please confirm you mean 40/40/20). Butterfly you wrote 70 for Sudan and 30 for children --- please give three integers that add to 100 in this exact order: Rainforest / Aid for Children / Sudan Humanitarian Aid. Once all three of you post explicit 3-number splits I will compute the combined proposal and suggest the next iterate. Keep it short.}
\userline{16:50:32}{Butterfly}{my allocation is 0/30/70}
\userline{16:50:35}{Swan}{humanitarian aid I think is most important than rainforest}
\facline{16:50:49}{Combined (rounded) average proposal: 13/35/52. Calculations: Rainforest (20+20+0)/3 $\approx$ 13.3 $\to$ 13; Children (35+40+30)/3 = 35; Sudan (45+40+70)/3 $\approx$ 51.7 $\to$ 52. Please reply with ``accept'' or give a new 3-number split. If you want suggest small tweaks ($\pm 5$) rather than large changes so we can converge quickly.}
\userline{16:50:53}{Llama}{50/25/25}
\facline{16:51:07}{New combined (rounded) average: 23/32/45 based on Llama 50/25/25, Swan 20/40/40, Butterfly 0/30/70. Please reply with ``accept'' or a small tweak ($\pm 5$ on any category) and a new 3-number split so we can converge quickly.}
\userline{16:51:23}{Llama}{accept}
\userline{16:51:32}{Swan}{go 40-40-20}
\userline{16:51:47}{Swan}{20 for rainforest}
\facline{16:52:18}{Combined proposal: 21/34/45. Calculation: (23+40+0)/3 = 21, (32+40+30)/3 = 34, (45+20+70)/3 = 45. Please reply ``accept'' or give a small tweak ($\pm 5$) as three integers summing to 100.}
\sysline{16:52:22}{The timer for this stage has ended; you can no longer respond.}
\sysline{16:52:28}{Butterfly has left the chat.}
\sysline{16:52:31}{Llama has left the chat.}
\sysline{16:52:31}{Swan has left the chat.}
\end{studyonebox}

\begin{studytwofacbox}{Study 2: Principles-Based Facilitator --- Invites Brief Reasoning of Values}
\userline{10:31:11}{Pig}{hey guys whats your allocation}
\userline{10:31:17}{Gorilla}{Hey guys I allocated 60 to woman and 20 to clean ocean and 20 to rainforest}
\userline{10:32:33}{Bear}{i went 45\% on woman 30\% on the ocean and 25\% on the forest but felt like i should change the forest allocation as they are also equally important}
\userline{10:32:41}{Pig}{40\% for Women 30\% to rainforest and 30\% to clean ocean}
\facline{10:32:49}{Thanks for sharing your initial allocations. What's one thing that influenced your own split or what do you notice about the allocations others have put forward?}
\userline{10:33:27}{Bear}{the importance of each organization globally}
\userline{10:33:43}{Gorilla}{true}
\facline{10:33:55}{It sounds like global importance is on your minds. Considering the different allocations shared does anyone want to respond to or build on an allocation idea from someone else?}
\userline{10:35:12}{Bear}{we could go equally but prioritize women and the forest}
\facline{10:35:35}{Bear that's an interesting principle. How do others feel about this idea of ``equally but prioritizing women and the forest'' especially as we work toward specific percentages across all three?}
\userline{10:35:50}{Gorilla}{I think woman should be top priority}
\userline{10:36:08}{Pig}{i think so too}
\sysline{10:36:12}{The timer for this stage has ended; you can no longer respond.}
\sysline{10:36:19}{Bear has left the chat.}
\sysline{10:36:27}{Gorilla has left the chat.}
\sysline{10:37:04}{Pig has left the chat.}
\end{studytwofacbox}

\begin{studytwofacbox}{Study 2: Summarization-Based Facilitator --- Summarizes Competing Priorities}
\facline{11:42:11}{Welcome everyone! Let's begin discussing the ideal allocation of donations across Global Housing for Orphans, Clean Ocean Action, and Rainforest Action. What are your initial thoughts on how we might split the funds?}
\userline{11:42:32}{Bear}{50\% on oceans 25 for orphans and 25 for rainforest}
\userline{11:43:11}{Dolphin}{I did 40\% rainforest 30\% orphans and oceans. I'm not particularly leaning towards any of the 3 since I think all are pretty important. So a little undecided I guess}
\userline{11:43:16}{Badger}{45\% to housing for orphans 25\% for clean ocean and 30\% for rainforest}
\facline{11:43:26}{Thank you all for sharing your initial ideas. We have proposals prioritizing Clean Ocean Action (Bear), Rainforest Action (Dolphin), and Global Housing for Orphans (Badger). What are some of the key factors or reasons influencing your initial splits?}
\userline{11:43:49}{Badger}{I would prioritize human beings than anything else}
\userline{11:44:10}{Bear}{understandable however oceans are very important for human beings too}
\userline{11:44:29}{Dolphin}{I allocated a little more for rainforest since there's much less of the rainforest left than ocean (i think). The other two I think are equally important}
\userline{11:44:31}{Badger}{how are oceans important to human beings}
\facline{11:44:39}{It seems we're discussing different priorities here: direct human welfare, broader environmental impact on human well-being, and the urgency of disappearing resources. Understanding these different influences is key. What are your thoughts on comparing these types of impacts?}
\userline{11:44:49}{Bear}{Rainforests... I have some experience with rainforest charities back in uni and they were not very good.}
\userline{11:45:23}{Bear}{Oceans are important for oxygen production from algae fish production water availability climate local weather}
\facline{11:45:29}{It sounds like we're weighing a few different aspects now: direct human impact, urgent environmental conservation, broader long-term environmental benefits for humans, and even considerations around the effectiveness or track record of different charities. How do these various factors align or conflict for you?}
\userline{11:45:48}{Dolphin}{yeah i did look at the description for the rainforest charity and it seems more focused on campaigning}
\userline{11:46:36}{Badger}{yes I agree the environment is also very important}
\userline{11:46:45}{Dolphin}{I don't mind increasing my allocation for ocean since they mention plastic removal from oceans}
\userline{11:46:51}{Dolphin}{so a bit more specific}
\userline{11:46:55}{Badger}{so how should we allocate}
\userline{11:46:57}{Bear}{Yeah oceans are doing some real job. Especially cleaning oceans from plastic so how about 45 for oceans 25 for forest}
\userline{11:47:02}{Bear}{30 for orphans}
\userline{11:47:08}{Dolphin}{sounds good}
\sysline{11:47:11}{The timer for this stage has ended; you can no longer respond.}
\sysline{11:47:17}{Bear has left the chat.}
\sysline{11:47:22}{Badger has left the chat.}
\sysline{11:47:27}{Dolphin has left the chat.}
\end{studytwofacbox}

\begin{studytwonofacbox}{Study 2: No Facilitator --- No Consensus (Timeout)}
\userline{09:35:22}{Zebra}{Hello! I think donating to clean ocean action is the most important out of the three}
\userline{09:35:45}{Zebra}{what does everyone else think?}
\userline{09:36:50}{Bear}{I would give the most to global fund for women as this charity concerns people and rest are about nature}
\userline{09:36:51}{Buffalo}{I do think that cleaning the oceans is the most urgent and impactful of the three.}
\userline{09:37:10}{Zebra}{why do you think it's not important?}
\userline{09:37:26}{Zebra}{sorry I meant important}
\userline{09:37:36}{Buffalo}{Helping woman is more important than helping animals so I support it.}
\userline{09:38:13}{Zebra}{I think if we don't support nature it will have a negative affect on humans therefore I think it's the most importan}
\userline{09:38:45}{Buffalo}{I would rank cleaning the oceans and helping women as the two most urgent with helping animals as third.}
\userline{09:39:14}{Zebra}{So should we put cleaning the oceans as most important?}
\userline{09:39:29}{Bear}{I would put the women as more important}
\userline{09:39:43}{Bear}{but let's allocate like 30\% toward oceans}
\userline{09:40:09}{Zebra}{what are your arguments for allocating so much towards women?}
\sysline{09:40:15}{The timer for this stage has ended; you can no longer respond.}
\sysline{09:40:18}{Zebra has left the chat.}
\sysline{09:40:21}{Bear has left the chat.}
\sysline{09:40:24}{Buffalo has left the chat.}
\end{studytwonofacbox}

\begin{studytwonofacbox}{Study 2: No Facilitator --- Participants Self-Organize}
\userline{11:29:02}{Lion}{Hello}
\userline{11:29:05}{Bear}{hello}
\userline{11:29:08}{Cat}{Hi}
\userline{11:29:23}{Lion}{So how should we allocate the donations?}
\userline{11:29:34}{Cat}{I think aids 4 children should get 50\%}
\userline{11:29:45}{Bear}{ill be honest i just split it 50/50 between the water aid and village children}
\userline{11:30:00}{Cat}{30\% 4 global fund for women}
\userline{11:30:23}{Cat}{Water aid comes last for me}
\userline{11:30:28}{Lion}{Ok so Bear it's 25\% for water and 25\% for children?}
\userline{11:30:56}{Bear}{nah as in 50\% each one}
\userline{11:31:11}{Cat}{Bear we have 3 charities}
\userline{11:31:14}{Bear}{the women one is too vague for me}
\userline{11:31:19}{Lion}{So you don't want to allocate any donations for women then?}
\userline{11:31:28}{Lion}{I guess that's an option too}
\userline{11:31:43}{Cat}{I think we should allocate for all}
\userline{11:31:43}{Bear}{i know but whats the money actually getting used for exactly is my only concern}
\userline{11:32:05}{Lion}{Hmm}
\userline{11:32:25}{Bear}{allocating some is fine i just think the other two are a bigger a priority}
\userline{11:32:28}{Lion}{I went 20\% for women}
\userline{11:32:38}{Cat}{yes the cause is brilliant and global with the children}
\userline{11:32:40}{Lion}{50\% for children 30\% for ocean}
\userline{11:32:54}{Bear}{yeah thats fine}
\userline{11:32:56}{Cat}{I can agree with Lion}
\userline{11:33:03}{Bear}{sounds good}
\userline{11:33:21}{Cat}{50\% for children 30\% for ocean and 20\% for women}
\userline{11:33:33}{Lion}{ok}
\userline{11:33:36}{Cat}{ok}
\userline{11:33:39}{Bear}{cool}
\sysline{11:34:07}{The timer for this stage has ended; you can no longer respond.}
\sysline{11:34:10}{Bear has left the chat.}
\sysline{11:34:12}{Lion has left the chat.}
\sysline{11:34:20}{Cat has left the chat.}
\end{studytwonofacbox}

\end{document}